\documentclass[10pt,a4paper,reqno]{amsart}
\usepackage{amsfonts,amsthm,latexsym,amsmath,amssymb,amscd,amsmath, epsf}

\newtheorem{theo+}           {Theorem}
\newtheorem{prop+}           {Proposition}
\newtheorem{coro+}           {Corollary}
\newtheorem{lemm+}           {Lemma}
\newtheorem{conjecture}      {Conjecture}
\theoremstyle{definition}
\newtheorem{defi+}           {Definition}
\newtheorem{problem}         {Problem}
\newtheorem{example}           {Example}

\theoremstyle{remark}
\newtheorem{rema+}           {Remark}

\newenvironment{theorem}{\begin{theo+}}{\end{theo+}}
\newenvironment{proposition}{\begin{prop+}}{\end{prop+}}
\newenvironment{corollary}{\begin{coro+}}{\end{coro+}}
\newenvironment{lemma}{\begin{lemm+}}{\end{lemm+}}
\newenvironment{remark}{\begin{rema+}}{\end{rema+}}
\newenvironment{definition}{\begin{defi+}}{\end{defi+}}

\newcommand{\al}{\alpha}
\newcommand{\be}{\beta}

\newcommand {\eps} {\epsilon}

\newcommand \dq {\mathfrak d(z)}

\newcommand{\la}{\lambda}
\newcommand{\La} {\Lambda}

\newcommand{\bC}{\mathbb C}
\newcommand{\bR}{\mathbb R}

\newcommand{\C}{\mathcal C}

\newcommand \M {\mathcal {M}}
\newcommand \E {\mathcal E}
\newcommand \PP {\mathcal P}
\newcommand \LL {\mathcal L}
\newcommand \fL {\mathfrak L}

\newcommand \fV {\mathfrak V}
\newcommand \bL {\mathbb L}

\def\newop#1{\expandafter\def\csname #1\endcsname{\mathop{\rm
#1}\nolimits}}

\newop{slc}
\newop{sign}
\newop{mdeg}

\begin{document}
          \numberwithin{equation}{section}

          \title[Algebro-geometric aspects of Heine-Stieltjes theory]
          {Algebro-geometric aspects of Heine-Stieltjes theory}

\author[B.~Shapiro]{Boris Shapiro}
\address{Department of Mathematics, Stockholm University, SE-106 91
Stockholm,
         Sweden}
\email{shapiro@math.su.se}

\dedicatory{Dedicated to Heinrich Eduard Heine and his 140 years old riddle} 

\date{\today}

\begin{abstract}
The goal of the paper is to develop a Heine-Stieltjes theory for univariate linear differential operators of higher order.   
Namely, for   a given linear ordinary differential operator $\dq=\sum_{i=1}^k
Q_i(z)\frac{d^i}{dz^i}$ with polynomial
coefficients set $r=max_{i=1,\ldots,k} (\deg Q_{i}(z)-i)$. If
$\dq$ satisfies the conditions:\; i) $r\ge 0$\; and\; ii) $\deg
Q_{k}(z)=k+r$\; we call it a {\it non-degenerate higher Lam\'e operator}.
Following the classical approach of E.~Heine and
T.~Stieltjes,  see \cite {He}, \cite {St} we study the multiparameter spectral problem of finding 
all  polynomials
$V(z)$ of degree at most $r$ such that  the equation:
$$
\dq S(z)+V(z)S(z)=0
$$
has for a given  positive integer $n$  a  polynomial solution $S(z)$ of degree $n$.
We show that under some mild non-degeneracy assumptions
 there exist  exactly ${n+r \choose n}$ such polynomials
$V_{n,i}(z)$ whose corresponding   eigenpolynomials  $S_{n,i}(z)$ are of degree $n$. We
        generalize a number of well-known results in this area and discuss occurring  degeneracies.

\end{abstract}

\maketitle

\tableofcontents

\section  {Introduction and main results}

The algebraic form of the classical Lam\'e equation, \cite {WW}, ch. 23, was introduced by Gabriel Lam\'e in 1830's in connection with the  separation of variables in the Laplace equation   in $\bR^l$ with respect to elliptic coordinates. It  has the form: 
\begin{equation}Ê\label{eq:standLame}
Q(z)\frac{d^2S}{dz^2}+\frac{1}{2} Q'(z)\frac{dS}{dz}+V(z)S=0,
\end{equation} 
where $Q_l(z)$ is a real  polynomial of degree $l$ with all real and distinct  roots, and $V(z)$ is a polynomial of degree at most $l-2$ whose choice depends on what type of solution to 
(\ref{eq:standLame}) we are looking for.  In the second half of the 19-th century several celebrated mathematicians including M.~B\^ocher, E.~Heine, F.~Klein, T.~Stieltjes studied the number and different properties of the so-called Lam\'e polynomials of a given degree and certain kind. (They are also called Lam\'e {\bf solutions} Êof a certain kind.)  Such solutions to  (\ref{eq:standLame})  exist for certain choices of $V(z)$ and are characterized by the property that their logarithmic derivative is a rational function. For a given $Q(z)$ of degree  $l\ge 2$ with simple roots there exist $2^l$ different kinds of Lam\'e polynomials depending on whether this solution  is smooth at a given root of $Q(z)$ or has there a square root singularity, see details in \cite {Poo} and \cite {WW}. (An excellent modern study of these questions can be found in \cite {Ma2}.) In what follows we will concentrate on the usual polynomial solutions of   (\ref{eq:standLame}) and its various modifications.  

A {\it generalized Lam\'e equation}, see \cite {WW} is the second
order differential equation given by
\begin{equation}
        Q_{2}(z)\frac
        {d^2S}{dz^2}+Q_{1}(z)\frac{dS}{dz}+V(z)S=0,
        \label{eq:comLame}
        \end{equation}
where $Q_{2}(z)$ is a complex polynomial of degree $l$ and $Q_{1}(z)$ is
a complex polynom of degree at most $l-1$. The special case $l=3$ is widely known as the Heun equation.

The next  fundamental proposition announced   in \cite {He} and provided there with not a quite satisfactory proof  was undoubtedly the starting point of the classical Heine-Stieltjes theory. 
\begin{theorem}[Heine]\label{th:Heine}
       If the coefficients of $Q_{2}(z)$ and $Q_{1}(z)$ are
       algebraically independent, i.e. they do not satisfy an algebraic
       equation with integer coefficients  then for any 
       integer $n> 0 $  there exists exactly ${n+l-2\choose n}$
polynomials $V(z)$ of degree exactly  $(l-2)$ such that the equation
(\ref{eq:comLame}) has and unique (up to a constant factor) polynomial solution $S$ of degree exactly
$n$.
       \end{theorem}
   
   \begin{remark}    
       Notice that throughout this paper we count polynomials $V(z)$ individually and polynomials $S(z)$   projectively, i.e. up to a constant factor.  
\end{remark}

  Later  on a physically important and directly related to the original (\ref{eq:standLame}) special case of (\ref{eq:comLame}) when  $Q_{2}(z)$ and
       $Q_{1}(z)$ have all real, simple and interlacing zeros and the same
       sign of the leading coefficients was  considered separately by T.~Stieltjes and his followers. 
       The equation can be then written as follows:
       \begin{equation}
        \prod_{i=1}^l(z-\al_{i})\frac
        {d^2S}{dz^2}+\sum_{j=1}^l\be_{j}\prod_{j\neq
        i}(z-\al_{i})\frac{dS}{dz}+V(z)S=0,
        \label{eq:genLame}
        \end{equation}
        with $\al_{1}<\al_{2}<\ldots<\al_{l}$ real and
$\be_{1},\ldots,\be_{l}$  positive. 
 In particular, the next proposition was proved. 

\begin{theorem}[Stieltjes-Van Vleck-B\^ocher
\cite{St},
\cite{VV}, \cite{Bo} and \cite{Sz}]
       Under the  assumptions of (\ref{eq:genLame}) and for any  integer $n> 0$
       \begin{enumerate}

      \item there exist
exactly ${n+l-2\choose n}$ distinct 
polynomials $V$ of degree $(l-2)$ such that the equation
(\ref{eq:genLame}) has a polynomial solution $S$ of degree exactly
$n$.

\item each root of each $V$ and $S$ is real and simple, and
belongs to the interval $(\al_{1},\al_{l})$.

\item  none of the roots of $S$ can coincide with some of
$\al_{i}$'s. Moreover, ${n+l-2\choose n}$ polynomials $S$ are in
1-1-correspondence
with ${n+l-2\choose n}$ possible ways to distribute $n$ points into the 
$(l-1)$ open intervals
$(\al_{1},\al_{2})$, $(\al_{2},\al_{3})$,\ldots,
$(\al_{l-1},\al_{l})$.
\end{enumerate}
\label{th:HSV}
\end{theorem}

The polynomials $V$ and the corresponding polynomial solutions
$S$ of the equation~(\ref{eq:comLame}) (or, equivalently, of 
(\ref{eq:genLame})) are called {\it Van Vleck} and {\it Stieltjes} (or {\it
Heine-Stieltjes}) polynomials resp.

\medskip 

The case when $\al_{i}$'s and/or $\be_{j}$'s are
complex is substantially less studied, see
\cite {MFMGO} and \cite {Mar}.  One nice result in this set-up is as
follows, see \cite {Po}.

\begin{theorem}[Polya] If in the notation of (\ref{eq:genLame}) all
$\al_{i}$'s are complex and all $\be_{j}$'s are  positive
that all the roots of each $V$ and $S$ belong to the
convex hull $Conv_{Q_2}$ of the set of roots $(\al_{1},\ldots,\al_{l})$ of $Q_2(z)$.
\label{th:Po}
\end{theorem}

\begin{rema+} The situation when   all  the residues $\be_j$ are negative (for example, $Q_1(z)=-Q'_2(z)$)  or have different signs seems to differ drastically from the latter case, see e.g. \cite {vAD} and \cite {EG}.  Ê
Further interesting results on the
distribution of
the zeros of Van Vleck and
Stieltjes polynomials under weaker assumptions on $\al_{i}$'s and
$\be_{j}$'s were obtained in \cite {Mar1}, \cite {Mar2},
    \cite {Al}, \cite {AZ}, \cite {Z}. \end{rema+}Ê

\medskip 
In the present article  we extend the above three fundamental results
on generalized Lam\'e equations of the second order to the case of  higher orders and/or 
complex coefficients.   
    Namely, consider  an arbitrary  linear ordinary differential operator
     \begin{equation}
     \dq=\sum_{i=1}^k
Q_i(z)\frac{d^i}{dz^i},
\label{eq:BasicOp}
\end{equation}
    with polynomial coefficients.
    The number
$r=max_{i=1,\ldots,k} (\deg Q_{i}(z)-i)$ will be called the {\it
Fuchs index} of $\dq$.   The operator $\dq$ is called a {\it higher 
Lam\'e operator} if its Fuchs index $r$ is non-negative. In the case of the vanishing Fuchs index 
$\dq$ is usually called {\em exactly solvable} in the physics literature, see \cite {Tu}. This case is also of the special interest in  connection with the classical Bochner-Krall problem in the theory of orthogonal polynomials. 

The operator $\dq$ is called  {\em non-degenerate}Ê if $\deg Q_{k}(z)=k+r$. Notice, that non-degeneracy of $\dq$ is a quite natural condition equivalent to the requirement that $\dq$ has either a regular or a regular singular point at $\infty$.

     \medskip
     Given a higher Lam\'e operator $\dq$ 
      consider the multiparameter  spectral
problem as follows. 

\noindent 
{\bf Problem.}Ê For each positive integer $n$ find   all polynomials 
$V(z)$ of degree at most $r$ such that  the equation
\begin{equation}
\dq S(z)+V(z)S(z)=0
\label{eq:1}
\end{equation}
has a  polynomial solution $S(z)$ of degree $n$.

\medskip
    Following the classical terminology we call (\ref{eq:1}) a 
    {\it higher Heine-Stieltjes spectral problem},  $V(z)$ is called a {\it
    higher Van Vleck
polynomial}, and the corresponding polynomial $S(z)$ is called a {\it higher Stieltjes polynomial}.  Below we will often skip mentioning `higher'. 

\begin{rema+} Obviously,  any differential operator (\ref{eq:BasicOp})
has either a non-negative or a negative Fuchs index. In the latter case  it can be easily transformed into the 
operator with a non-negative Fuchs index by  the change of variable $y=\frac 1 z$.  Notice also that the condition of non-degeneracy is generically satisfied. 
 In what follows we will always assume wlog that the leading coefficient  of such an operator is  a monic polynomial.
     \end{rema+}

\subsection {Generalizations of Heine's theorem, degeneracies and nonresonance condition} 
     
   We start with  a number of generalizations of Heine's theorem~\ref{th:Heine}.     
   Following Heine's original proof one can obtain the following straightforward generalization.

      \begin{theorem}
     For any non-degenerate higher Lam\'e operator $\dq$ with algebraically
independent coefficients of its polynomial coefficients $Q_{i}(z),\,i=1,\ldots,k$ and for any   
$n\ge 0$ there exist
exactly ${n+r \choose r}$ distinct Van Vleck polynomials
$V(z)$'s whose corresponding   Stieltjes polynomials $S(z)$'s 
are unique (up to a constant factor) and of degree $n$. 
   \label{th:higHei} 
        \end{theorem}

       Our next result obtained by a linear-algebraic interpretation of (\ref{eq:1}) has no genericity assumptions and is crucial in   the problem  of  existence of solutions  of  (\ref{eq:1}), comp.   \cite{Mar}, Problem 1. 

\begin{theorem}\label{th:main} 
For any non-degenerate operator $\dq$ with a Fuchs index $r\ge 0$ and any  positive integer $n$ the total number  of   Van Vleck polynomials $V(z)$ (counted with natural multiplicities) having a Stieltjes polynomial  $S(z)$  of degree less than or equal to  $n$ equals $\binom {n+r+1}{r+1}$.  
\end{theorem}

 \begin{rema+}Ê Note that in Theorem~\ref{th:main}  we do not require that there is a unique (up to constant factor) Stieltjes polynomial corresponding to a given Van Vleck polynomial. For the exact description of the notion  of the {\em natural multiplicity} of a Van Vleck polynomial which is rather lengthy consult Definition~\ref{def:multi} in \S~\ref{sec:proofs}. 
  \end{rema+}

  
\noindent  
{\bf On degeneracies.}Ê Notice that   Theorem~\ref{th:higHei} claims that  a generic operator $\dq$ has for any positive $n$ exactly $\binom {n+r}{r}$ distinct Van Vleck polynomials each of which has a unique Stieltjes polynomial and this polynomial is of degree exactly $n$. 
  The question about possible degeneracies occurring Ê in Problems~(\ref{eq:comLame}) and (\ref{eq:1})  if we drop the genericity assumptions on $\dq$ is quite delicate.  Not only Van Vleck polynomials can attain a nontrivial multiplicity as well as more than $1$-dimensional linear space of Stieltjes polynomials but there are examples when there are no Stieltjes polynomials of some degree.  In particular,  for any polynomial $Q(z)$ of degree $l$ no choice of a polynomial $V(z)$ of degree at most $l-2$ will supply the equation  
       $$    Q(z)\frac {d^2S}{dz^2}-Q'(z)\frac{dS}{dz}+V(z)S=0$$
        with a polynomial solution $S$ of degree $l+1$.  (This follows from the Proposition 5 and Lemma 4 of \cite{EG}.) 
        The fact that  (\ref{eq:comLame}) can admit families (linear spaces of dimension at least $2$) of polynomial solutions $S$ corresponding to one and the same $V$ was already   mentioned by Heine in his original proof.  More exact information is available nowadays.  For example,  a result of Varchenko-Scherbak gives necessary and sufficient condition for a Fuchsian second order equation to have 2 independent polynomial solutions, see \cite{SchV} and \cite{EG}.     Finally, high multiplicity of Van Vleck polynomials occur, for example, in the case $Q_2(z)=z^l,\; Q_1(z)=0$. Then one can easily show that for all $n\ge 2$ there exists just one and only polynomial $V_n(z)=-n(n-1)z^{l-2}$ solving the above problem; its corresponding Stieltjes polynomial equals $S_n(z)=z^n$. The multiplicity of the latter Van Vleck polynomial $V_n(z)$ is 
        $\binom {n+l-2}{n}$.

To formulate necessary and sufficient conditions under which the conclusion of Heine's theorem holds for all positive integers $n$  is apparently an impossible task.   Heine himself mentions that for the validity of his result for a given  fixed positive integer $n$ one has to avoid a certain discriminantal hypersurface (similar to the usual discriminant of  univariate polynomials) which is given by an equation with integer coefficients but this  equation is difficult to obtain explicitly. 

Below we formulate  a simple sufficient condition which allows us to avoid many of the above degeneracies and guarantees the existence of Stieltjes polynomials of a given degree. Namely, 
  consider  an arbitrary  non-degenerate operator $\dq  $ of the form (\ref{eq:BasicOp}) with  the Fuchs index $r$.    Denote by  $A_k, A_{k-1},..., A_1$ the  coefficients at the highest possible degrees $k+r, k+r-1,..., r+1$ in the polynomials $Q_k(z), Q_{k-1}(z),...,Q_1(z)$ resp. (Notice that any subset of $A_j$'s can vanish but $A_k\neq 0$ due to the non-degeneracy of $\dq$.)  In what follows we will often use  the notation 
  $$(j)_i=j(j-1)(j-2)...(j-i+1),$$
  where $j$ is a non-negative and $i$ is a positive integer. In case $j=i$ one has $(j)_i=j!$ and in case $j<i$ one gets $(j)_i=0$.  For any non-negative $n$ we call  by {\it the $n$-th diagonal coefficient} $\bL_n$ Êthe expression: 
  \begin{equation}\label{eq:leading}
  \bL_n=(n)_kA_k+(n)_{k-1}A_{k-1}+....+(n)_1A_1.
  \end{equation}


\begin{proposition}\label{pr:nonres} 
If in the above notation and for a given positive integer $n$   the  {\bf  $n$-th nonresonance  condition} 
   \begin{equation}Ê  \label{eq:nonres}Ê    
   \bL_n\neq \bL_j, \; j=0,1,...,n-1   \end{equation}   holds 
then there exist Van Vleck polynomials which possess  Stieltjes polynomials of degree exactly $n$ and no other Stieltjes polynomials of degree smaller than $n$. In this case the total number of such Van Vleck polynomials (counted with natural multiplicities) equals $\binom {n+r}{r}$. 
 \end{proposition} 
 
 \begin{rema+}Ê The above  nonresonance condition is quite natural. It says that if the equation (\ref{eq:1}) has a polynomial solution of degree $n$ then it has no polynomial solutions of smaller degrees. 
Another way to express this fact is  that  if the indicial equation of (\ref{eq:1}) at $\infty$ has $-n$ as its root then it has no roots among non-positive integers $0, -1, -2, ..., 1-n$, see e.g. \cite{Poo}, ch. V. 
 \end{rema+} 
 
 Explicit formula (\ref{eq:leading}) for $\bL_n$ immediately shows that  Theorem~\ref{th:main}  and Proposition~\ref{pr:nonres} are valid  for any non-degenerate $\dq$ and all sufficiently large $n$. 

\begin{corollary}
    For any non-degenerate higher Lam\'e operator $\dq$ and all sufficiently large $n$ the $n$-th   nonresonance  condition holds. In particular,  for any problem (\ref{eq:1}) there exist and finitely many (up to a scalar multiple) Stieltjes polynomials of any sufficiently large degree.

        \label{cor:NewhigHei} 
        \end{corollary}
        
   \begin{rema+}ÊNotice that for an arbitrary non-degenerate operator $\dq$ and a given integer $n$ it is difficult to find explicitly  all Van Vleck polynomials which possess a Stieltjes  polynomial of degree at most $n$. By this we mean that in order to do this one has, in general, to solve  an overdetermined system of algebraic  equations in the coefficients of  $V$ since the set of Van Vleck polynomials under consideration is not a complete intersection. (This system of determinantal equations  contains many more equations than variables.) However, one  consequence of Heine's way to prove his Theorem~\ref{th:Heine} is as follows.  For a given non-degenerate operator $\dq$ with Fuchs index $r$ and a positive  integer $n$  denote by $\fV_n\subset Pol_r$ the set of all its  Van Vleck polynomials possessing a Stieltjes polynomial of degree exactly $n$. 
    \end{rema+} 

\begin{theorem}\label{th:HeineAdd} If in the above notation the $n$-th  nonresonance condition (\ref{eq:nonres}) holds Ê  then $\fV_n$ is a complete intersection and the corresponding system of equations can be given explicitly in each specific case. 
\end{theorem} 

Explicit example of the defining system of $r$ algebraic equations in $r$ variables can be found in \S~\ref{sec:proofs}, see Example~\ref{ex:1}. 
In  purely linear algebraic setting this result and further information about relevant discriminants can be found in \cite {SS}. 

     


\subsection{Generalizations of Stieltjes's theorem} 

We continue with a conceptually new generalization of Theorem~\ref{th:HSV}. It was conjectured by the present author after extensive computer experiments and was later proved by  P.~Br\"anden. In the present paper we only announce  his result  and its corollaries since it requires a large amount  of additional information and techniques. The actual proof will be published by its author elsewhere.   

\begin{definition}Ê
A differential operator $\dq=\sum_{i=m}^kQ_i(z)\frac{d^i}{dz^i},\;1\le m\le k$ where all $Q_i(z)$'s are polynomials with real coefficients is called a {\em strict hyperbolicity preserver}  if for any real polynomial $P(z)$ with all real and simple roots the image $\frak d(P(z))$  either vanishes identically or is  a polynomial with only real and simple roots.
\end{definition}Ê

\begin{theorem}\label{th:higStie} For any strict hyperbolicity preserving non-degenerate Lam\'e operator $\dq$ with the Fuchs index $r$ as above  and any integer  $n\ge m$  
  \begin{enumerate}

      \item there exist
exactly ${n+r\choose n}$ distinct 
polynomials $V(z)$ of degree exactly $r$  such that the equation
(\ref{eq:1}) has a polynomial solution $S(z)$ of degree exactly
$n$.

\item all roots of each such $V(z)$ and $S(z)$ are real, simple, coprime.

\item ${n+r\choose n}$ polynomials $S(z)$ are in 1-1-correspondence
with ${n+r\choose n}$ possible arrangements of $r$ real roots of
 polynomials V(z) and $n$ real roots of the corresponding polynomials $S(z)$. 
\end{enumerate} 
\end{theorem}Ê

Using Theorem~\ref{th:main}  one immediately sees that Êthe latter result describes  the set of all possible pairs $(V,S)$ with $m\le n=\deg S$ for any hyperbolicity preserver $\dq$. 

\begin{rema+} The interested reader can check that the sum of the first two terms in (\ref{eq:genLame}) is indeed a strict hyperbolicity preserver. It looks very tempting and important  to find an analog of the electrostatic interpretation of the roots of classical Heine-Stieltjes and classical Van Vleck polynomials  (alias 'Bethe ansatz') in the case of higher Heine-Stieltjes and  Van Vleck polynomials, comp. \cite{MMM}.  
\end{rema+}Ê 

\begin{rema+} Notice that the converse to the above theorem is false. Namely, one can show that the exactly solvable operator $\dq(f) = f' + z(z+1)f''$ has all hyperbolic  eigenpolynomials but is not a hyperbolicity preserver. 
\end{rema+}

A straight-forward application of Theorem~\ref{th:higStie} to differential operators of order $2$  gives the  following. 
Consider a differential equation
\begin{equation}
        \prod_{i=1}^l(z-\al_{i})\frac
        {d^kS}{dz^k}+\sum_{j=1}^l\be_{j}\prod_{j\neq
        i}(z-\al_{i})\frac{d^{k-1}S}{dz^{k-1}}+V(z)S=0,
        \label{eq:higLame}
        \end{equation}
where $2\le k\le l$, $\al_{1}<\al_{2}<\ldots<\al_{l}$ and
$\be_{1},\ldots,\be_{l}$ are positive. 

\begin{corollary}
       Under the  assumptions of (\ref{eq:higLame}) and for any $n\ge k-1$ 
       \begin{enumerate}

      \item there exist
exactly ${n+l-k\choose n}$
polynomials $V(z)$ of degree $(l-k)$ such that the equation
(\ref{eq:higLame}) has a polynomial solution $S(z)$ of degree exactly
$n$.

\item all roots of each $V(z)$ and $S(z)$ are real, simple, coprime and
belong to the interval $(\al_{1},\al_{l})$.

\item ${n+l-k\choose n}$ polynomials $S(z)$ are in 1-1-correspondence
with ${n+l-k\choose n}$ possible arrangements of $(l-k)$ real roots of
a polynomial $V(z)$ and $n$ real roots of the corresponding polynomial $S(z)$
on the interval
$(\al_{1},\al_{l})$.
\end{enumerate}
\label{th:higHSV}
\end{corollary}

It seems that Theorem~\ref{th:higStie}Ê and Corollary~\ref{th:higHSV} give a new 
interpretation
of Theorem~\ref{th:HSV}  even in the classical case (\ref{eq:genLame}).
However  the following two statements proven by G.~Shah 
 explain this mystery, see Theorem 3
of \cite{Sh1} and Theorem 3 of \cite {Sh3}.

\begin{prop+}
   Under the assumptions of Theorem~\ref{th:HSV} the
     roots of any Van Vleck polynomial $V(z)$ and its corresponding
     $S(z)$ are coprime. \end{prop+}

     Moreover,

\begin{prop+} If $v_{1}<v_{2}<\ldots<v_{r},\;r=l-2$ denote the roots of some
Van Vleck polynomial $V(z)$ in the classical situation~(\ref{eq:genLame}) then
for each $i=2,\ldots, l-1$ the interval
$(v_{i}, \al_{i+1})$ contains no roots of the corresponding $S(z)$.  Therefore, for each polynomial $S(z)$ the distribution of its
$n$ roots into $(l-1)$ intervals $(\al_{1},\al_{2}),\ldots,
(\al_{l-1},\al_{l})$ coincides with the distribution of these roots
defined by the roots of its Van Vleck polynomial $V(z)$.
      \end{prop+}

\begin{rema+} Note
that we do not claim $v_{i}< \al_{i+1}$, i.e. the endpoints of the
interval $(v_{i}, \al_{i+1})$ can be placed in the wrong order or can
coincide.
\end{rema+}

\subsection{Generalizations of Polya's theorem}Ê

 We start with a simple-minded statement  of Polya's theorem~\ref{th:Po}, \cite{Po}. 

\begin{theorem}
       If the zeros $\al_{1},\ldots,\al_{l}$ in (\ref{eq:higLame}) are
       complex and the constants $\be_{1},\ldots,\be_{l}$ are
       non-negative then all the roots of $V$'s and $S$'s lie in the (closed)
       convex hull $Conv_{Q_k}$ of the roots $(\al_{1},\ldots,\al_{l})$  of the polynomial $Q_k(z)$.
\label{th:higPol}
\end{theorem}

The next  Theorem~\ref{th:locl} is far more  general.  It shows that Theorem~\ref{th:higPol}  is {\bf asymptotically} true for any non-degenerate Lam\'e operator. The question for which operators $\dq$ the roots of all its Van Vleck and Stieltjes polynomials lie exactly (and not just asymptotically) in the convex hull of its leading coefficient seems to be very hard even in the classical case of the equation (\ref{eq:comLame}).

\medskip    
    
\begin{theorem} 

 For any non-degenerate higher Lam\'e operator $\dq$ and any $\eps>0$ there exists a positive integer $N_\eps$ such that the 
     zeros of all Van Vleck polynomials $V(z)$ possessing   a Heine-Stieltjes
     polynomial $S(z)$ of degree $n\ge N_\eps$ and well as all zeros of these Stieltjes polynomials belong to $Conv_{Q_{k}}^{\eps}$.
     Here $Conv_{Q_{k}}$ is the convex hull of all zeros of the leading coefficient $Q_{k}$ 
     and $Conv_{Q_{k}}^{\eps}$ is its $\eps$-neighborhood in the usual Euclidean distance on $\bC$. 
        
    \label{th:locl}
    \end{theorem}
    

The latter theorem is closely related to  the next somewhat simpler localization result having  independent interest. 

\begin{proposition}\label{pr:local}
For any non-degenerate higher Lam\'e operator $\dq$ there exist a positive integer $N_0$ and a positive number $R_0$ such that all zeros  of all Van Vleck polynomials $V(z)$ possessing a  Stieltjes polynomial $S(z)$ of degree $n\ge N_0$ as well as all zeros of these Stieltjes polynomials  lie in the disk $\vert z\vert \le R_0$.  
\end{proposition} 

\begin{remark} Notice that the roots of absolutely all Van Vleck polynomials (and not just those whose Stieltjes polynomials are of sufficiently large degree) of any  non-degenerate higher Lam\'e operator $\dq$ lie in some disk. But this is no longer true for Stieltjes polynomials. If the set of all Stieltjes polynomials is discrete (up to a scalar multiple) then their roots are bounded. But as soon as some Van Vleck polynomial admits an at least $2$-dimensional linear space of Stieltjes polynomials then these roots become unbounded for obvious reasons. However for sufficiently large $n$ no Van Vleck polynomial admits such families, see Corollary~\ref{cor:NewhigHei} and the localization result holds. 

\end{remark}

\begin{remark}
Similar and  stronger localization results with explicit constants and degree bounds  were independently obtained by J.~Borcea (private communication).  
\end{remark}
    
 Let us now show a typical behavior of the zeros of Van Vleck polynomials and the corresponding Stieltjes polynomials obtained in numerical experiments. Below we consider as an example the operator 
 $\dq=Q(z)\frac {d^3}{dz^3}$ with  $Q(z)=(z^2+1)(z-3I-2)(z+2I-3).$ For $n=24$ we calculate  all 25 
 pairs $(V,S)$ with  $\deg S=24$. (Notice that $V$ in this case is linear.)   The asymptotic behavior of the union of zeros of all Van Vleck polynomials whose Stieltjes polynomials have a given degree $n$ when $n\to \infty$ as well as the asymptotics of the zeros of subsequences of Stieltjes polynomials  of increasing degrees whose corresponding (monic) Van Vleck polynomials  have a limit seems to be  an extremely rich and interesting topic, see first steps in \cite {ShT}.

\begin{figure}[!htb]
\centerline{\hbox{\epsfysize=3.0cm\epsfbox{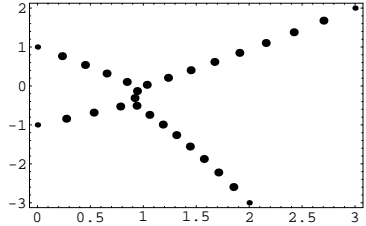}}}

\vskip 1cm

\caption{Zeros of $25$ different linear Van Vleck polynomials whose Stieltjes polynomials are of degree $24$.   Four average size dots are the zeros of $Q(z)=(z^2+1)(z-3I-2)(z+2I-3)$ 
and  large dots are the zeros of different $V(z)$.}
\label{fig1}
\end{figure}


\begin{figure}[!htb]
\centerline{\hbox{\epsfysize=1.5cm\epsfbox{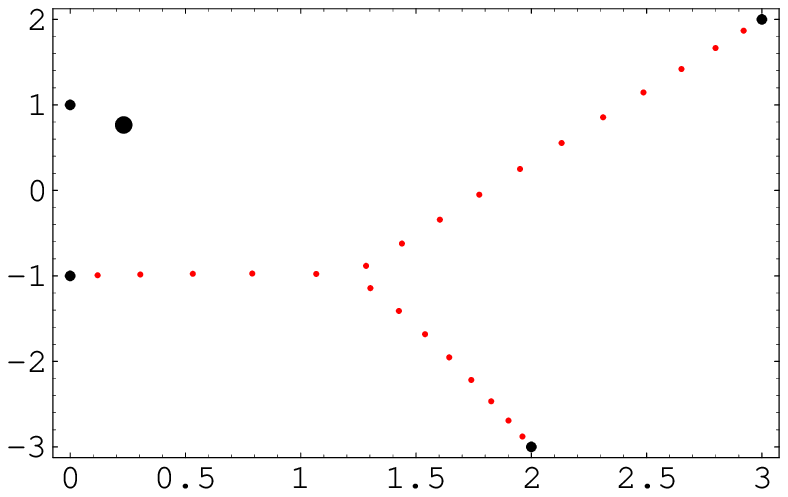}}
\hskip0.5cm\hbox{\epsfysize=1.5cm\epsfbox{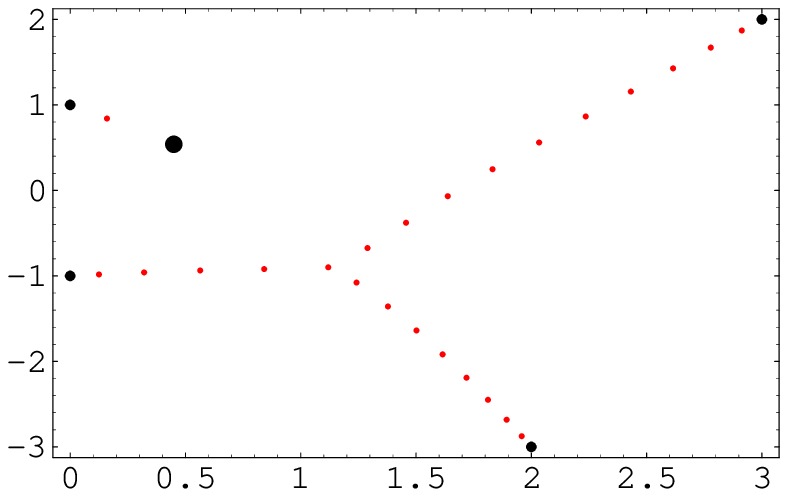}}
\hskip0.5cm\hbox{\epsfysize=1.5cm\epsfbox{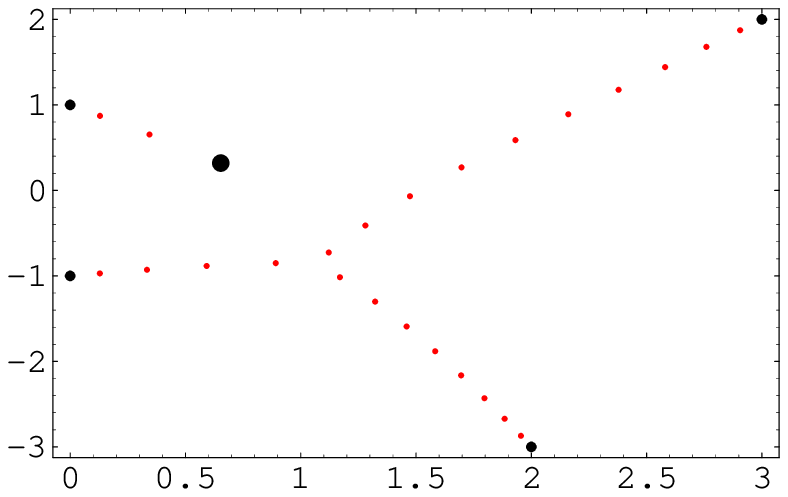}}
\hskip0.5cm\hbox{\epsfysize=1.5cm\epsfbox{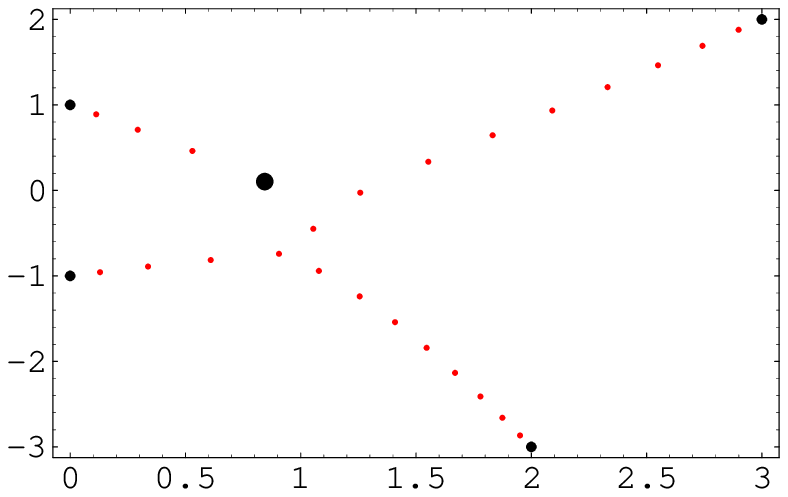}}
\hskip0.5cm\hbox{\epsfysize=1.5cm\epsfbox{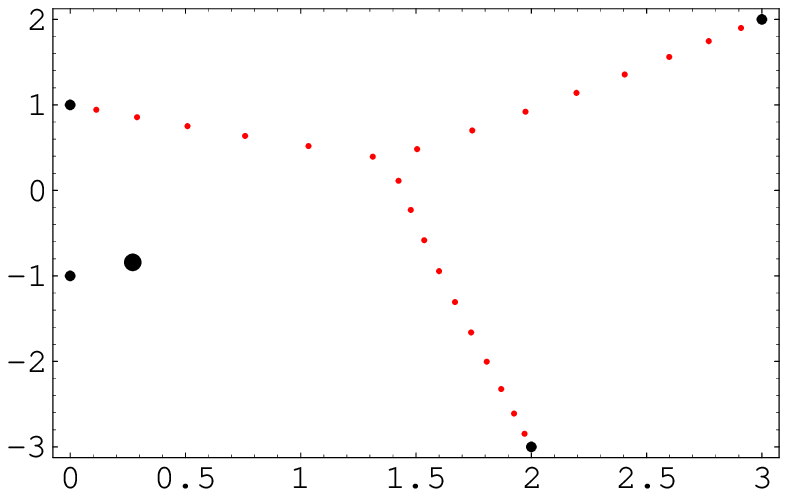}}
}

\centerline{\hbox{\epsfysize=1.5cm\epsfbox{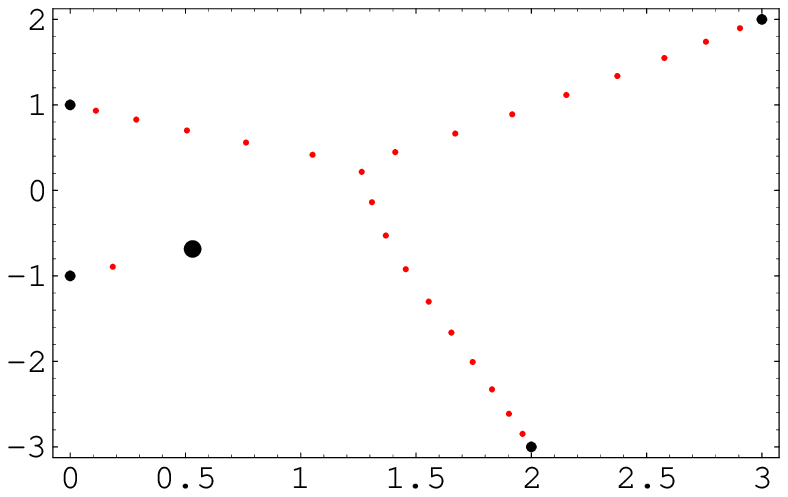}}
\hskip0.5cm\hbox{\epsfysize=1.5cm\epsfbox{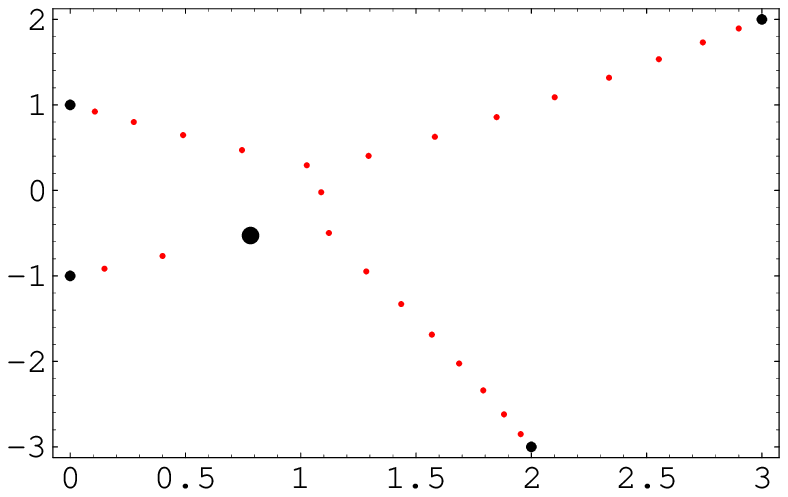}}
\hskip0.5cm\hbox{\epsfysize=1.5cm\epsfbox{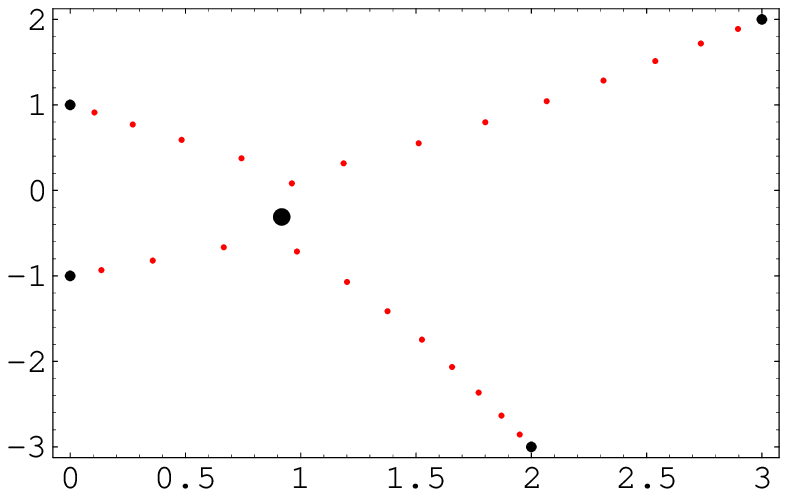}}
\hskip0.5cm\hbox{\epsfysize=1.5cm\epsfbox{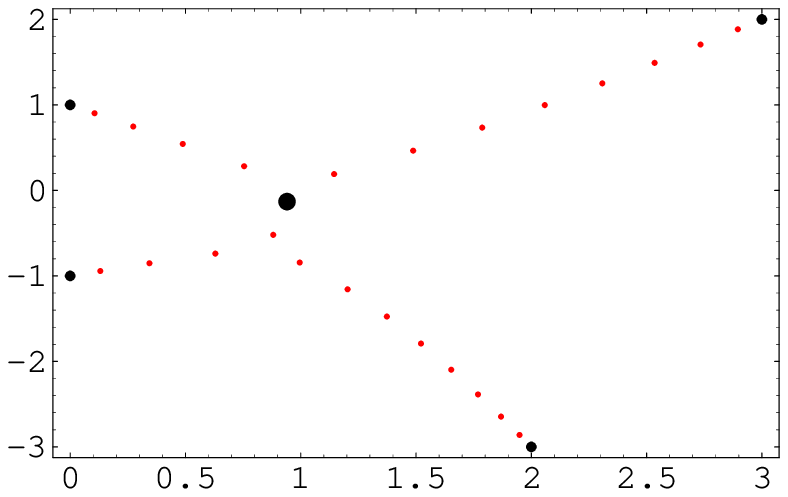}}
\hskip0.5cm\hbox{\epsfysize=1.5cm\epsfbox{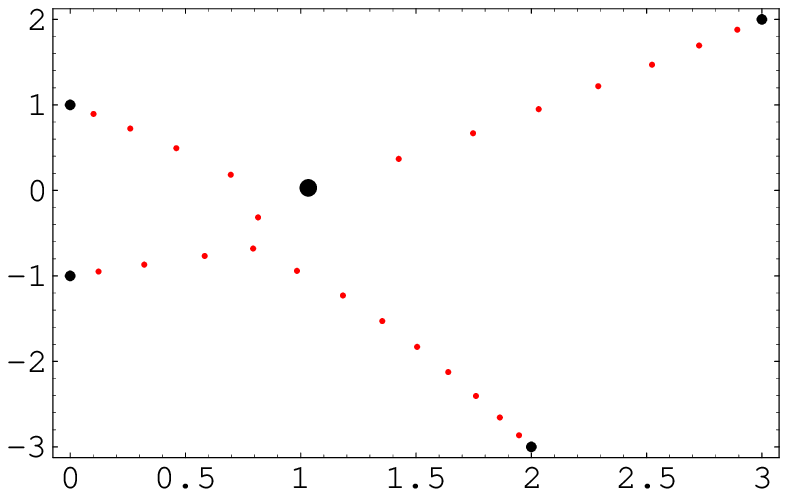}}
}

\centerline{\hbox{\epsfysize=1.5cm\epsfbox{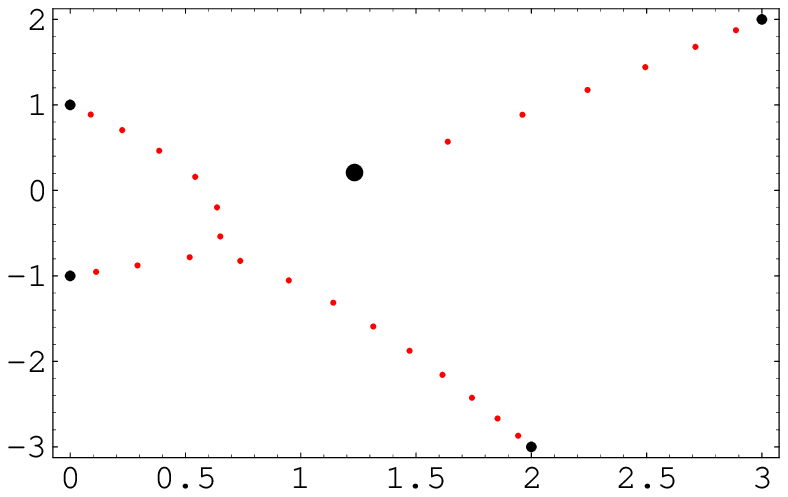}}
\hskip0.5cm\hbox{\epsfysize=1.5cm\epsfbox{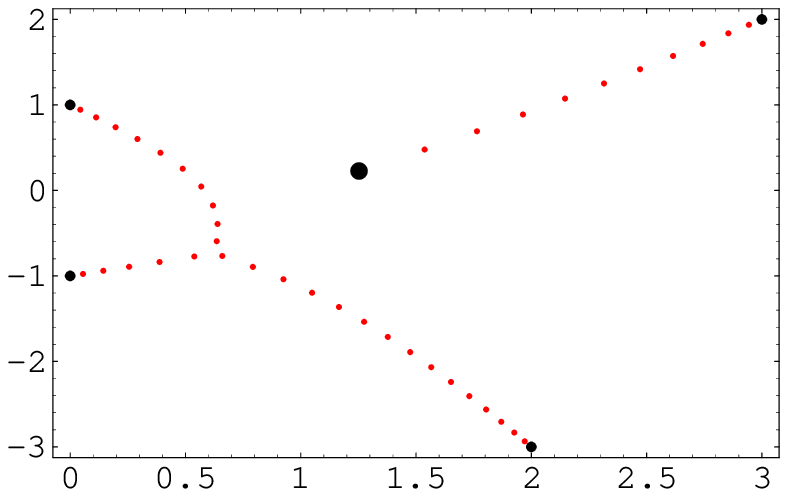}}
\hskip0.5cm\hbox{\epsfysize=1.5cm\epsfbox{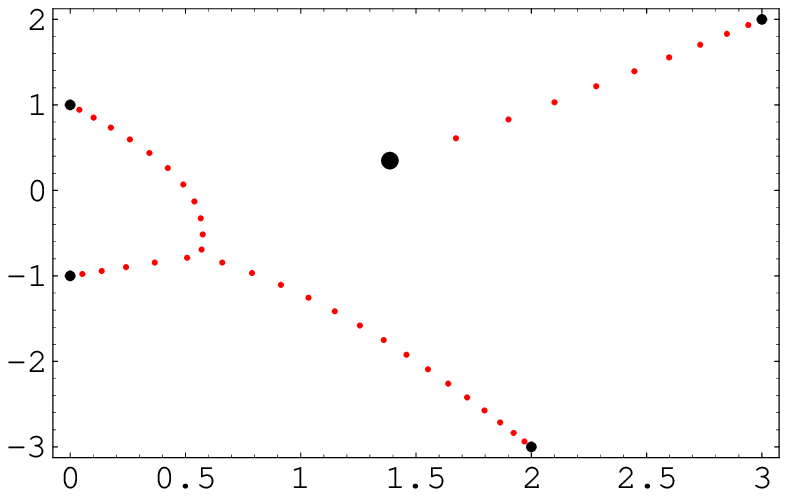}}
\hskip0.5cm\hbox{\epsfysize=1.5cm\epsfbox{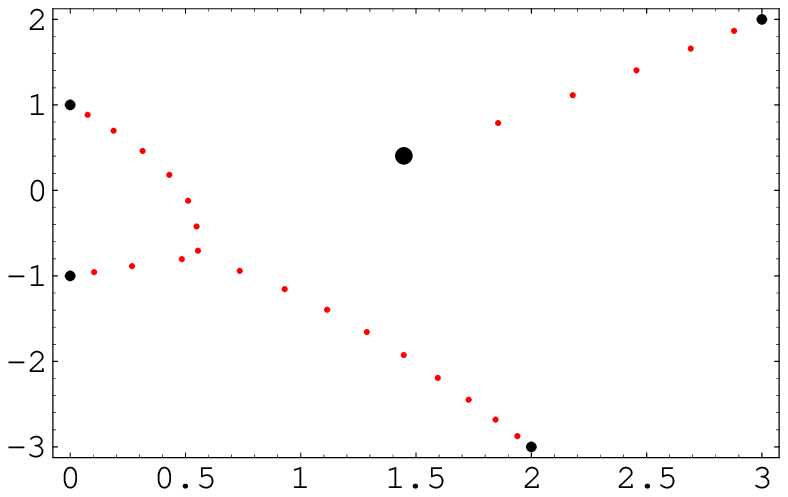}}
\hskip0.5cm\hbox{\epsfysize=1.5cm\epsfbox{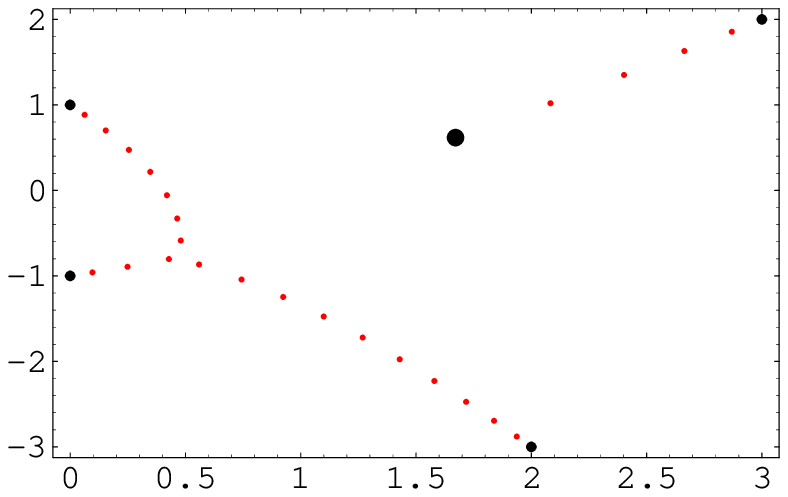}}
}

\centerline{\hbox{\epsfysize=1.5cm\epsfbox{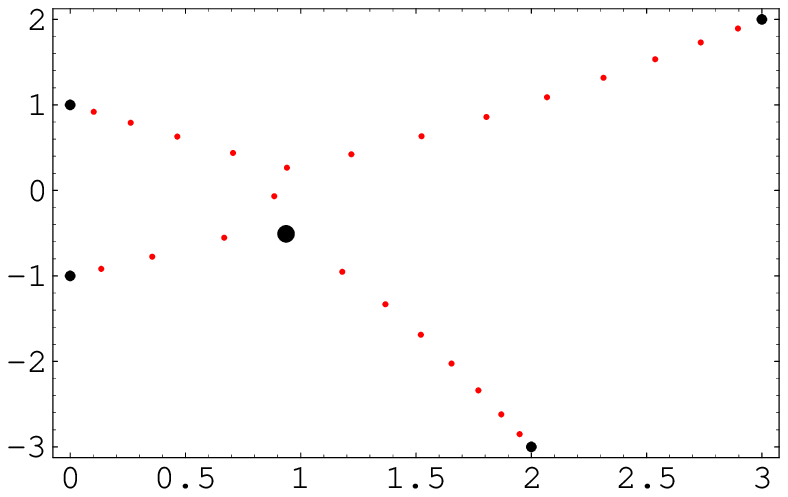}}
\hskip0.5cm\hbox{\epsfysize=1.5cm\epsfbox{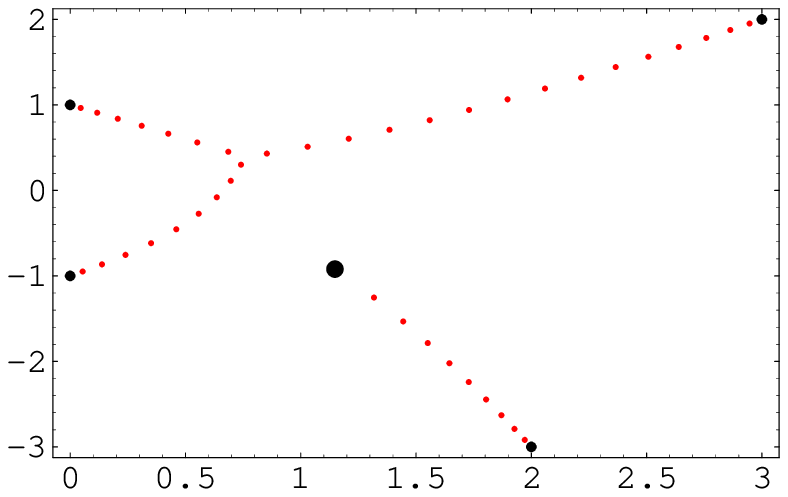}}
\hskip0.5cm\hbox{\epsfysize=1.5cm\epsfbox{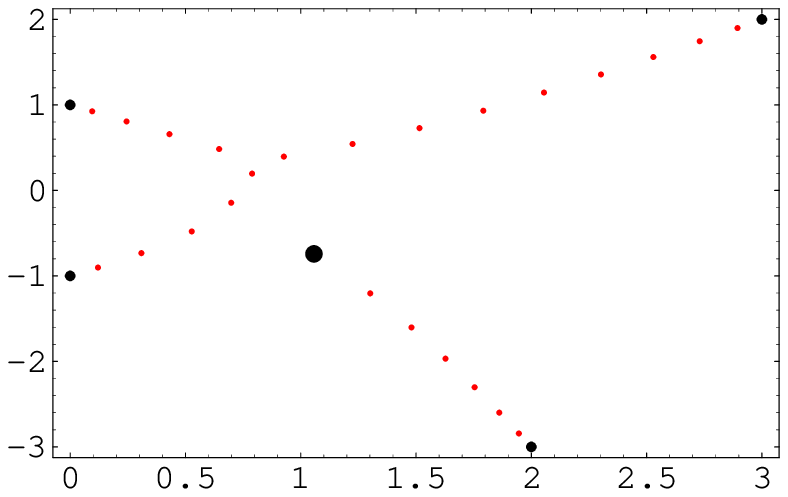}}
\hskip0.5cm\hbox{\epsfysize=1.5cm\epsfbox{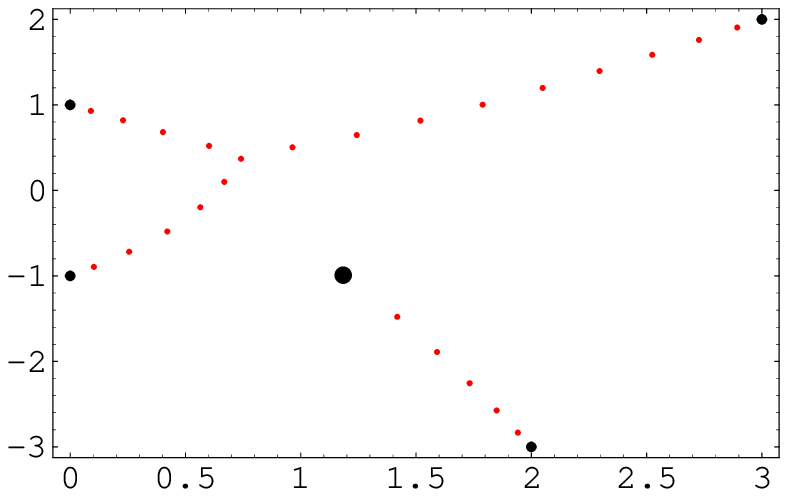}}
\hskip0.5cm\hbox{\epsfysize=1.5cm\epsfbox{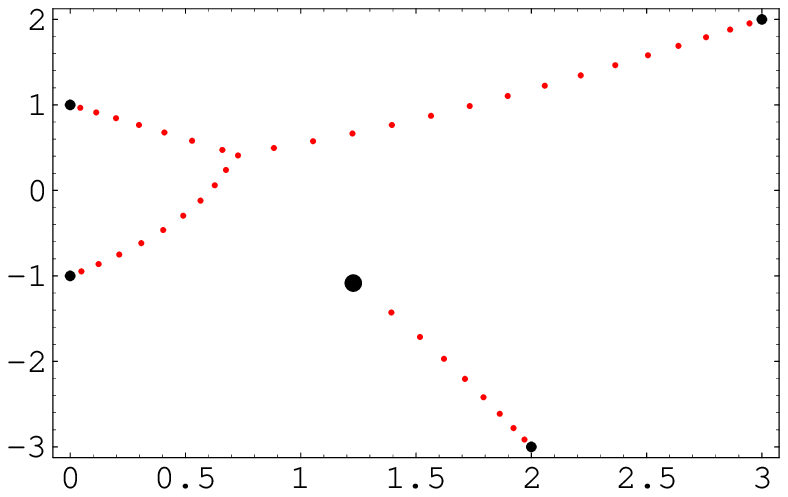}}
}

\centerline{\hbox{\epsfysize=1.5cm\epsfbox{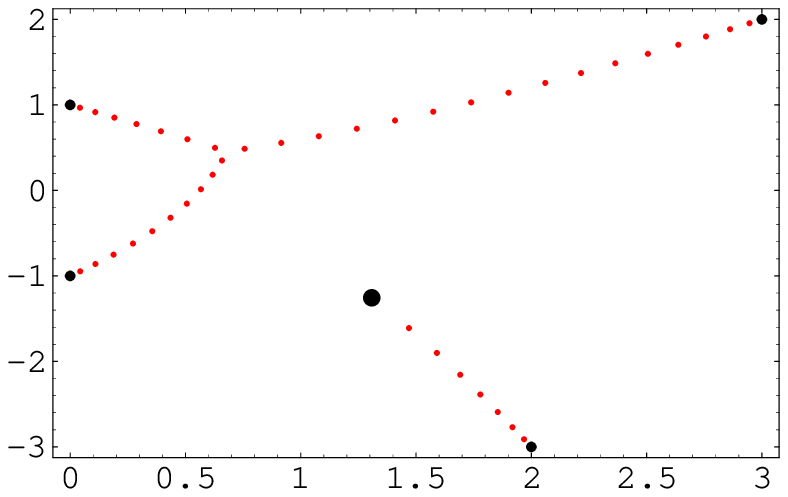}}
\hskip0.5cm\hbox{\epsfysize=1.5cm\epsfbox{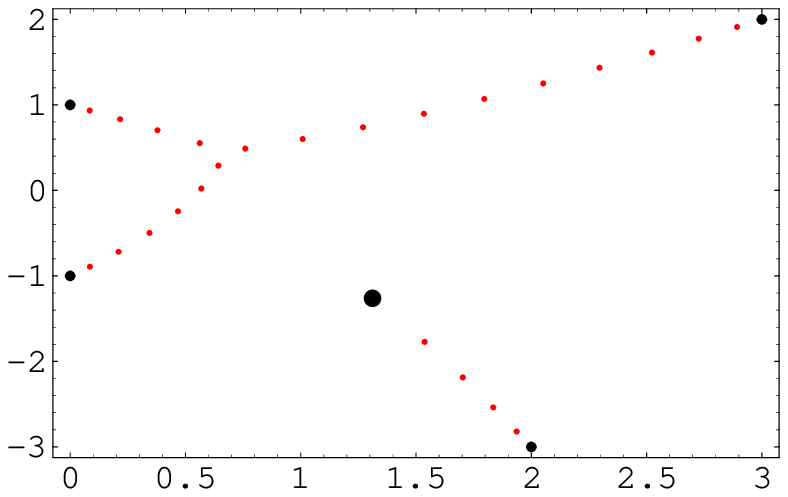}}
\hskip0.5cm\hbox{\epsfysize=1.5cm\epsfbox{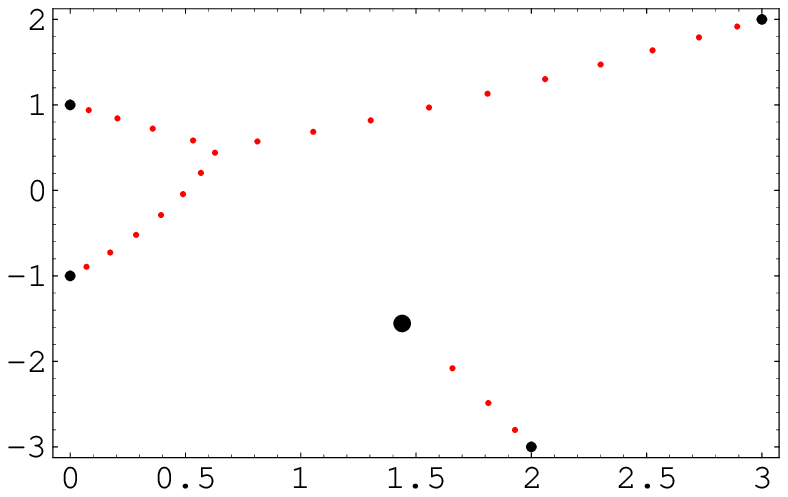}}
\hskip0.5cm\hbox{\epsfysize=1.5cm\epsfbox{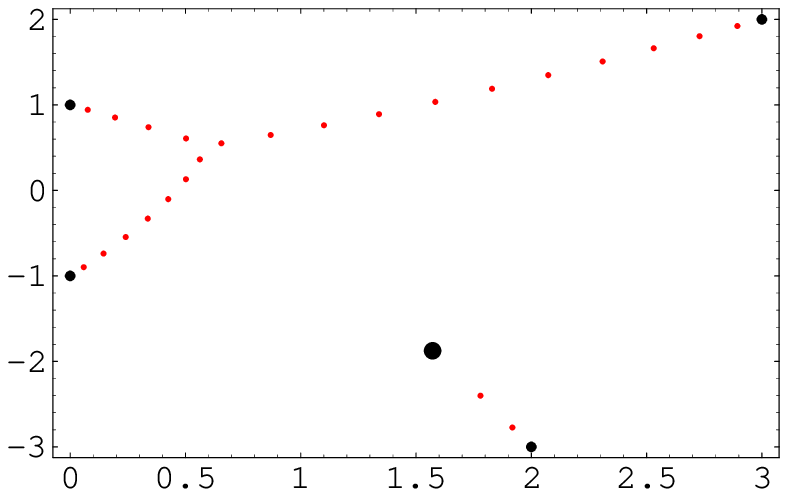}}
\hskip0.5cm\hbox{\epsfysize=1.5cm\epsfbox{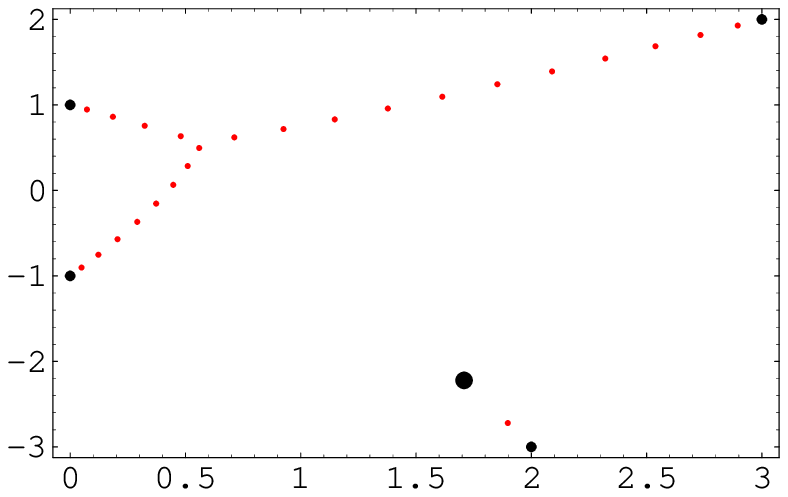}}
}

\vskip 1cm

\caption{Zeros of $25$ different Stieltjes polynomials of degree $24$
for the above $\dq$.  The small dots on each picture are the
$24$ zeros of $S(z)$; $4$ average size dots are the zeros of $Q(z)$ 
and the single large
dot is the (only) zero of the corresponding $V(z)$.}
\label{fig2}
\end{figure}

\medskip

{\em Some literature.}Ê Let us mention a few relatively recent  references on (generalized) Lam\'e equation. 
 Being an object 
of substantial 
physical and mathematical importance it,  in particular,  gives an example of an equation whose  monodromy group can be analyzed in details, see \cite {BW}, \cite {Ma1}-\cite {Ma2}. It is also closely
related to the so-called quasi-exact solvability and integrable models, see \cite{Etin}.  Theory  of multiparameter spectral problems originating
from Heine-Stieltjes pioneering studies was developped in sixtees, see
e.g. \cite {Vol} and references therein.  Recently the interest to Heine-Stieltjes
polynomials has stimulated by an unexpected extension of the Bethe ansatz in
representation theory, see \cite {ReV}, \cite{MuV}, \cite{Sch}, \cite {SchV} and a further series of article by A.~Varchenko and his coauthors, e.g.  \cite {MuTV}.  Starting with  \cite {MFS} a substantial  progress has been made in the
understanding of the asymptotics of the root distributions for these
polynomials  when either $l \to\infty$ (thermodynamic asymptotics) or $n \to \infty$ (semi-classical
asymptotics), see \cite {Bou}, and \cite {BJOT}. 
Asymptotic root distribution for  the eigenpolynomials of non-degenerate exactly solvable operators was studied in \cite{MS} and \cite{BR}.  Interesting preliminary results of a similar flavor in the case of degenerate exactly solvable operators were very recently obtained by T.~Bergkvist, \cite {Ber}. 


{\em Acknoledgements.} I am very  grateful to my former 
collaborator and colleague G.~M\'asson
for the pioneering numerical experiments in 1999. His wild guesses
       gave birth to a vast project on asymptotics for polynomial
       solutions to linear ordinary differential equations depending on
       parameter(s) which occupies me since then.  I want to thank  P.~Br\"anden for sharing his proof of Theorem~\ref{th:higStie} with me in a private communication. Sincere thanks go  to 
       R.~ B\o gvad, J.~Borcea,  I.~Scherbak, A.~Varchenko  and, especially, to A.~Mart\'inez-Finkelshtein for many useful  discussions of the area and their interest in my work.  
       Finally, I owe  a great deal to
       the Wolfram corporation whose  package Mathematica although quite 
expensive and  not quite reliable was indispensable  in  doing the actual
       $\mu\alpha\theta\eta\mu\alpha\tau\iota\kappa{\alpha}$.

         \section {Proof of generalized Heine's theorems}
	 \label{sec:proofs}
	 
	We start with Theorem~\ref{th:higHei} (see Introduction). For this we need a  detailed description of  the action of a non-degenerate operator $\dq$  on the linear space $Pol_n$ of all univariate polynomials of degree at most $n$. 

           \begin{proof} 
           Substituting $V(z)=v_{r}z^r+v_{r-1}z^{r-1}+\ldots+v_{0}$ and $ S(z)=s_nz^n+s_{n-1}z^{n-1}+\ldots+s_{0}$ in (\ref{eq:1}) we get the following system of  $(n+r+1)$ equations of a band shape (i.e.  only a fixed  and independent of  $n$ number of diagonals is non-vanishing  in this system): 
	     \begin{equation} 
	     \begin{cases}\label{eq:system} 
	     0=Ês_n(v_r+L_{n,n+r});\\
	     0=s_n(v_{r-1}+L_{n,n+r-1})+s_{n-1}(v_r+L_{n-1,n+r-1});\\
	     0=s_n(v_{r-2}+L_{n,n+r-2})+s_{n-1}(v_{r-1}+L_{n-1,n+r-2}) +s_{n-2}(v_r+L_{n-2,n+r-2});\\
	    \vdots  \qquad  \qquad  \vdots  \qquad \qquad  \vdots \qquad  \qquad \vdots  \qquad \qquad \vdots \qquad  \qquad \vdots\qquad  \qquad \vdots\\
	   0=s_n(v_0+L_{n,n})+s_{n-1}(v_{1}+L_{n-1,n})+...+ s_{n-r}(v_r+L_{n-r,n});\\
	      0=s_n L_{n,n-1}+s_{n-1}(v_{0}+L_{n-1,n-1})+     ...+s_{n-r-1}(v_r+L_{n-r-1,n-1});\\
	      0=s_n L_{n,n-2}+s_1L_{1,r+1}+s_2(v_{r}+L_{2,r})+...+s_{n-r-2}(v_r+L_{n-r-2,n-2});\\
	       \vdots  \qquad  \qquad  \vdots  \qquad \qquad  \vdots \qquad  \qquad \vdots  \qquad \qquad \vdots \qquad  \qquad \vdots\qquad  \qquad \vdots\\
              0=s_{n} L_{n,r}+s_{n-1}L_{n-1,r}+s_{n-2}L_{n-2,r}+...+s_{0}(v_r+L_{0,r});\\
	     0=s_{n}L_{n,r-1}+s_{n-1}L_{n-1,r-1}+s_{n-2}L_{n-2,r-1}+...+s_{0}(v_{r-1}+L_{0,1});\\ 
	      \vdots  \qquad  \qquad  \vdots  \qquad \qquad  \vdots \qquad  \qquad \vdots  \qquad \qquad \vdots \qquad  \qquad \vdots\qquad  \qquad \vdots\\   
	      0=s_{n} L_{n,1} +s_{n-1}L_{n-1,1}+...+s_{1}(v_{0}+L_{1,1})+s_{0}(v_{1}+L_{0,1});\\ 
	       0=s_{n} L_{n,0} +s_{n-1}L_{n-1,0}+...+s_{1}L_{0,1}+s_{0}(v_0+L_{0,0}).
	     \end{cases}
	     \end{equation}
		     
	    Here $L_{p,q}$ is a polynomial  which expresses  the coefficient containing  $s_p$ at the power $z^q$ in $\sum_{i=1}^kQ_i(z)S^{(i)}$.  Obviously, it is linear  in the coefficients of $Q_k(z),...,Q_1(z)$  and is  explicitly given by the relation 
	    $$L_{p,q}=\sum_{r=1}^k(p)_rA_{r,q-p+r},$$
	    where $A_{r,q-p+r}$ is the coefficient at $z^{q-p+r}$ in $Q_r(z)$.  In the notation used in the definition (\ref{eq:leading}) we have $L_{m,m+r}=\bL_m, m=0,...,n$. 
	    We use the convention that  $L_{p,q}$ vanishes outside the admissible range of indices and, therefore many of the above coefficients $L_{p,q}$ are in fact  equal to $0$.    
	    (In the system (\ref{eq:system}) we assumed that $n\ge r$ for simplicity.) 
	    Notice that all equations in  (\ref{eq:system}) depend linearly on the variables $v_{r},...,v_0$ and $s_{n},...,s_{0}$ as well as on the
	     coefficients of polynomials $Q_{i}(z),\;i=1,\ldots,k.$ Note
	      additionally, that (\ref{eq:system}) is lower-triangular w.r.t the
	     coefficients $s_{n},...,s_0$ which allows us to perform  the following important elimination.  Let us
	     enumerate the equations of (\ref{eq:system}) from $0$ to $n+r$ assigning the number
	     $j$ to the equation describing the vanishing of the coefficient
	     at the power $z^{n+r-j}$. Then if $\bL_n=L_{n,n+r}\neq 0$ one has that the $0$-th equation has a solution  
	     $s_n=1$ and $v_r=-L_{n,n+r}\neq 0$.  
The next $n$
	     equations are triangular w.r.t the
	     coefficients $s_{n},...,s_0$, i.e.
	     $j$-th equation in this group  contains only  the variables $s_{n},s_{n-1},\ldots,s_{n-j}$ (among all $s_j$'s) along  with other types
	     of variables. Thus    under the assumption
	     that all the diagonal terms $v_r+L_{n-i,n+r-i}=\bL_{n-i}-\bL_n,\;i=0,1,...,n$ are nonvanishing  we can express all   $s_{n-i},\;i=0,1,...,n$ consecutively as rational functions of  the remaining variables and get the reduced system of
	     $r$  rational equations containing only  
	     $(v_{r-1},\ldots,v_{0})$ as unknowns.  Notice that in view of $v_r=-L_{n,n+r}\neq 0$ the non-vanishing of the diagonal entries $v_r+L_{n-i,n+r-i},\;i=0,1,...,n$ coincides exactly with the nonresonance  condition (\ref{eq:nonres}). 

	    Cleaning the common denominators we get a  reduced system of polynomial equations. We show now that this 
	     polynomial system  is quasi-homogeneous in the variables 
	     $v_{j}$ with the quasi-homogeneous weights $w(v_{j})$ given by $w(v_{j})=r-j$.
	     Thus using the weighted-homogeneous version of the Bezout theorem, see e.g
	     \cite {Do} we get that  if the system under
	     consideration defines a complete intersection, i.e. has only isolated solutions then their number  (counted with
	     multiplicities)  equals  ${n+r\choose r}$ in the corresponding weighted projective space. To check the quasi-homogenuity note that the standard 
	     action of $\bC^{*}$ on the set of roots of the polynomial $V(z)$ by simultaneous multiplication assigns the weight
               $r-j$ to its coefficient $v_{j}$. These weights are still valid in the reduced system with the  variables   $s_n,...,s_0$ eliminated.  Finally, we have to show that if the coefficients of $Q_k(z),...,Q_1(z)$ are algebraically independent then the eliminated system has exactly ${n+r\choose r}$ simple solutions. Indeed,  
               consider the linear  space $EQ$  of all systems of $r$ quasi-homogeneous equations in the variables $(v_r,...,v_0)$ with the weights $w(v_{j})=r-j$ and where the   $i$-th equation is weighted-homogeneous of degree $n+i$.  We equip this space with the standard monomial basis.

To accomplish the proof of  Theorem \ref{th:higHei} we need two additional standard facts.

               \begin{lemma}\label{lm:discr} ÊThe discriminant $Discr\subset EQ$  (i.e. the set of the coefficients  of monomials in the equations for which the system has at least one solution of multiplicity greater than $1$) is given by an algebraic equation with rational coefficients in the  standard monomial basis of $EQ$.
               \end{lemma} 
               \begin{proof} See \cite {GKZ}, ch. 13. 
               \end{proof} 
               
   Consider some linear parameter  space $\La$ with a chosen  basis. Assume there is a rational map: 
   $\Phi:\La\to EQ$ where each coordinate  in the standard monomial basis   of $EQ$  is given by a rational function with rational coefficients w.r.t to the chosen basis in $\La$.  The next statement is obvious. 
   
   \begin{lemma}\label{lm:pullback} 
   In the above notation the pullback of $\Phi^{-1}(Discr)$ in $\La$ either a) coincides with the whole 
   $\La$ or b) is given in the chosen basis by an algebraic equation with rational coefficients.  
   \end{lemma}Ê

    It remains to show that there are some values of the coefficients of the polynomials $Q_k(z),...,Q_1(z)$ for which there are exactly ${n+r\choose r}$ distinct solutions of (\ref{eq:1}).  Here we are not able to follow the nice inductive argument of \cite{He}, see also the last paragraph in \S~\ref{sec:heine}.  Heine's proof does not generalize immediately to higher order equations. Instead we can, for example,  invoke Theorem~\ref{th:higStie}  whose proof is completely independent of the present arguments.  
    It claims, in particular, that for any strict hyperbolicity preserver of the form $\dq=\sum_{i=m}^kQ_i(z)\frac{d^i}{dz^i}$ and any $n\ge m$ there exist exactly $n+r \choose r$ pairs $(V,S)$. One can additionally choose such a hyperbolicity preserver with $m=1$ and therefore get a necessary example of an operator with given $k$  and $r$ such that for any $n\ge 1$ it has exactly the maximal number of pairs $(V,S)$. 
\end{proof}

To settle Theorem~\ref{th:main} (see Introduction) let us first reinterpret  Problem (\ref{eq:1}) in linear algebraic terms.

\subsection {On eigenvalues for rectangular matrices}

We start with the following natural question. 

\medskip 
\noindent 
{\bf Problem.} Given a $(l+1)$-tuple of $(m_1\times m_2)$-matrices $A, B_1,..., B_l$ where $m_1\le m_2$ describe the set of all values of parameters $\la_1,...\la_l$ for which the rank of the linear combination $A+\la_1B_1+...+\la_lB_l$ is less than $m_1$  i.e. when the linear system   $v*(A+\la_1B_1+...+\la_lB_l)=0$Ê  has a nontrivial (left) solution $v\neq 0$ which we call an {\em eigenvector of $A$ wrt the linear span of $B_1,...,B_l$}.
\medskip 

Let  $\M_{m_1,m_2}$ denote the linear space  of all $(m_1\times m_2)$-matrices with complex entries. 
Below we will consider $l$-tuples of $(m_1\times m_2)$-matrices $B_1,..., B_l$ which are linearly independent in $\M_{m_1,m_2}$ and denote their linear span by $\LL=\LL(B_1,...,B_l)$. 
Given a matrix pencil   $\PP=A+\LL$ where $A\in \M_{m_1,m_2}$ denote by $\E_\PP\subset \PP$ its {\it eigenvalue locus}, i.e. the set of matrices in $\PP$ whose rank is less than the maximal one.   Denote by $\M^1\subset \M_{m_1,m_2}$ the set of all $(m_1\times m_2)$ matrices with positive corank, i.e whose rank is less than $m_1$.  
Its co-dimension equals $m_2-m_1+1$ and its degree as an algebraic variety equals $\binom {m_2}{m_1-1}$,  see 
\cite {BV}, Prop. 2.15.  Consider the natural left-right action of the group $GL_{m_1}\times GL_{m_2}$ on $\M_{m_1,m_2}$, where  $GL_{m_1}$ (resp. $GL_{m_2}$) acts on $(m_1\times m_2)$-matrices by the left (resp. right) multiplication. This action on  $\M_{m_1,m_2}$ has finitely many orbits, each orbit being the set of all matrices of a given (co)rank, see e.g. \cite{AVG}, ch.1 \S 2. Notice that due to  the well-known  formula of the product of coranks the codimension of the set of matrices of rank $\le r$ equals $(m_1-r)(m_2-r)$.  Obviously, for any pencil $\PP$ one has that the eigenvalue locus coincides with $\E_\PP=\M^1\cap \PP$. Thus for a generic pencil $\PP$ of dimension $l$ the eigenvalue locus $\E_\PP$ is a subvariety of $\PP$ of codimension $m_2-m_1+1$ if $l\ge m_2-m_1+1$ and it is empty otherwise. The most interesting situation for applications occurs  when $l=m_2-m_1+1$ in which case $\E_\PP$ is generically a finite set. From now on  let us assume that $l=m_2-m_1+1$. Denoting as above  by $\LL$ the linear span of 
$B_1,...,B_{l}$ we say that $\LL$ is {\it transversal to $\M^1$} if the intersection $\LL\cap \M^1$ is finite and {\it non-transversal to $\M^1$}Ê  otherwise.  Notice that due to homogeneity of $\M^1$ any  $(m_2-m_1+1)$-dimensional linear subspace $\LL$ transversal to it   intersects $\M^1$ only at $0$ and that the multiplicity of this intersection at $0$  equals $\binom{m_2}{m_1-1}$. 

We start with the  following obvious  statement which will later imply Theorem~\ref{th:main}.

\begin{lemma}\label{lm:int}
If $(m_2-m_1+1)$-dimensional linear space $\LL$ is tranversal to $\M^1$ then for any matrix $A\in \M_{m_1,m_2}$ the eigenvalue locus $\E_\PP$ of the pencil $\PP=A+\LL$ consists of exactly $\binom {m_2}{m_1-1}$ points counted with multiplicitites. 
\end{lemma}


\begin{rema+} Notice that since $\M^1\subset \M(m_1,m_2)$ is an incomplete intersection then  in order to explicitly determine the eigenvalue locus of a given matrix $A$ w.r.t. some $(m_2-m_1+1)$-dimensional linear subspace  $\LL\subset \M(m_1,m_2)$ one has to solve an overdetermined system of $\binom{m_2}{m_1}$  equations describing the vanishing of all maximal minors of a $(m_1\times m_2)$-matrix depending on parameters.  
 Fortunately, in our main application, i.e. for the multi-parameter spectral problem (\ref{eq:1}) we encounter   only  'triangular' rectangular matrices (i.e. with the left-lower corner vanishing) for which the determination of the eigenvalue locus often reduces to a complete intersection, see proof of Theorem~\ref{th:higHei} and Theorem~\ref{th:HeineAdd}. 
\end{rema+} 

Let us explain how Lemma~\ref{lm:int}Ê implies Theorem~\ref{th:main}. ÊNamely,   given a non-degenerate operator  $\dq$ in order to find all its Van Vleck polynomials having (at least one)  Stieltjes polynomial of degree at most $n$ we need to study  the action of $\dq$ on the linear  space $Pol_n$ of all univariate polynomials of degree at most $n$. If $\dq$ has the Fuchs index $r$ then $\dq$ maps $Pol_n$ to $Pol_{n+r}$. 
Using the standard monomial basis $1, z, z^2,..., z^l$ in $Pol_l$ we get that if $n\ge k=\text{ord}(\dq)$ then   the action of $\dq$ in this basis is represented by a 'triangular' band $(n+1)\times(n+r+1)$-matrix $A_{\dq,n}$ with at most  $r+k$ non-vanishing diagonals.  Here 'triangular' means that all entries $a_{i,j}$ of $A_{\dq,n}$ with $i<j$ vanish.   Denote by $I_s,\; s=0,...,r$ the $(n+1)\times (n+r+1)$-matrix whose entries are given by $a_{i,j}=0$ if $i-j\neq s$ and $1$ otherwise. 
Denote by $\fL$  the linear span  of $I_0,...,I_{r}$ and notice that $\fL$ is transversal to $\M^1\subset M_{n+1,n+r+1}$  since any matrix belonging to the pencil  $\fL$ and different from $0$ has full rank. 

Notice that adding  an arbitrary polynomial $V(x)=v_rz^r+v_{r-1}z^{r-1}+...+v_0$ of degree at most $r$  to $\dq$ corresponds on the matrix level to adding  of the linear combination  $v_rI_0+v_{r-1}I_1+...+v_0I_r$ to the initial matrix $A_{\dq,n}$. The existence of a non-trivial Stieltjes polynomial of degree at most $n$ corresponds to the fact that the  matrix  $A_{\dq,n}+v_rI_0+v_{r-1}I_1+...+v_0I_r$ has a non-trivial (left) kernel. Thus, for a given non-degenerate operator $\dq$ the problem of finding  all Van Vleck polynomials whose Stieltjes polynomials are of degree at most $n$ is exactly equivalent to the determination of all the eigenvalues  of its matrix $A_{\dq,n}$ w.r.t. the linear space $\fL$ in the above-mentioned sense.  Lemma~\ref{lm:int} has a simple analog for 'triangular' rectangular matrices which is equivalent to Theorem~\ref{th:main}.  

Namely, denote by $TM(m_1,m_2)\subset\M(m_1,m_2), \quad m_1\le m_2$ Êthe set of all 'triangular' $m_1\times m_2$-matrices, i.e. with $a_{i,j}=0$ for $i<j$.  Let $\fL\subset 
TM(m_1,m_2)$ be the linear subspace spanned by all $(m_2-m_1+1)$ possible unit matrices $I_1,...,I_{m_2-m_1+1}\in TM(m_1,m_2)$. Finally, denote by $TM^1\subset TM(m_1,m_2)$ the set of all 'triangular' matrices with positive corank. 

\begin{lemma}\label{lm:fl} For any matrix $A\in TM(m_1,m_2)$ the eigenvalue locus $\E_\PP$ of the pencil $\PP=A+\fL$ consists of exactly $\binom {m_2}{m_1-1}$ points counted with multiplicitites. 
\end{lemma} 

\begin{proof}ÊThe same as above. 
\end{proof}Ê
 
The latter Lemma settles Theorem~\ref{th:main}. Let us now prove Proposition~\ref{pr:nonres} (see Introduction). 

\begin{proof} In the above notation consider the pencil $\mathcal A_n(v_r,v_{r-1},...,v_0)=A_{\dq,n}+v_rI_0+v_{r-1}I_1+...+v_0I_r$ of 
$(n+1)\times(n+r+1)$-matrices.  One has 
$$\mathcal A_n(v_r,v_{r-1},...,v_0)=\begin{pmatrix}Ê\bL_n+v_r &*&*&*&*&\cdots\\
                              0          &\bL_{n-1}+v_r&*&*&*&\cdots\\
                              0        &0   &\bL_{n-2}+v_r&*&*&\cdots\\
                              0        &0  &0  &\bL_{n-3}+v_r&*&\cdots\\
                              \vdots &\vdots &\vdots &\vdots&\vdots&\vdots
                              \end{pmatrix},
$$  
where $*$ stands for possibly non-vanishing entries.  
The obvious necessary condition   for such a matrix to have a positive corank it that one of the elements on the shown above main diagonal vanishes, i.e. there exists $i=0,...,n$ such that $\bL_i+v_r=0$ or, equivalently,  $v_r=-\bL_i$.   Set $v_r=-\bL_n$ thus 'killing' the entry in the left-upper corner. Recall that  the $n$-th nonresonance condition requires that $\bL_n\neq \bL_j, j=0,...,n-1$. Therefore the subtraction of $\bL_n$ along the main diagonal will keep 
all other diagonal entries except for the left-upper corner non-vanishing.   The $r$-dimensional pencil $\mathcal A_n(-\bL_n,v_{r-1},...,v_0)=A_{\dq,n}-\bL_nI_0+v_{r-1}I_1+...+v_0I_r$  has its $1$-st column of the above matrix presentation vanishing and its main diagonal being the same  for all possible values of $v_{r-1},...,v_0$. Therefore, by Lemma~\ref{lm:int}Ê there exist exactly $\binom {n+r} {r}$  eigenvalues  $v_{r-1},v_{r-2},...,v_0$ counted with multiplicities such that   $\mathcal A_n(-\bL_n,v_{r-1},...,v_0)$   has a positive corank. Finally, notice that since for  any matrix from  the pencil $\mathcal A_n(-\bL_n,v_{r-1},...,v_0)$ its entries along the  main diagonal except for the left-upper corner  are non-vanishing its corank  can be at most $1$. 
Moreover, when the corank of such a matrix is $1$ then the occurring non-trivial linear combination of the rows which vanishes must necessarily include the first row since the second, the third etc rows are linearly independent for the above reason. 
The coefficients of this linear dependence of rows are exactly  the coefficients of  the corresponding Stieltjes polynomial. The fact that  the first row must be in the linear dependence  means in this language that the leading coefficient of this Stieltjes polynomial (which by definition is of degree at most $n$)  must be non-vanishing, i.e. this Stieltjes polynomial is of degree exactly $n$.   Proposition~~\ref{pr:nonres} is settled.   
   \end{proof} 

To prove  Theorem~\ref{th:HeineAdd}Ê (see Introduction) we need to take a more careful  look   at the proof of Theorem~\ref{th:higHei}. Namely,  consider again the system (\ref{eq:system}) 
determining the set of all pairs $(V,S)$  where $V$ is a Van Vleck polynomial and $S$ is the corresponding Stieltjes polynomial of degree at most $n$. As in the proof of Theorem~\ref{th:higHei}  
we  solve the $0$-th equation in  (\ref{eq:system})  by taking $s_n=1$ and $v_r=-L_{n,n+r}=-\bL_n$. 
(By Proposition~\ref{pr:nonres} Êfinding a solution of  (\ref{eq:system}) with   $s_n=1$ and $v_r=-L_{n,n+r}=-\bL_n$ leads to a pair $(V,S)$ Êsuch that $V$ is of degree exactly equal to $r$  and $S$ is of degree exactly equal to $n$.) Then we express consecutively the variables $s_{n-1}, s_{n-2},...,s_0$ from the next $n$ equations of (\ref{eq:system}). The crucial circumstance here is that while doing this we  only  divide by the differences of the form $\bL_n-\bL_j,\; j=n-1,n-2,...,0$ which are non-vanishing 
due to the validity of our nonresonance condition. Substituting  the obtained expressions for $s_j,\;j=0,...,n$ in the remaining $r$ equations in  (\ref{eq:system}) we get the required eliminated system of algebraic 
equations on the variables $v_{r-1},...,v_0$ which proves the required result. \qed

To illustrate the above procedure let us consider a concrete example. 

\begin{example}\label{ex:1} Consider the action of some operator $\dq$ with the Fuchs index $r=2$ on the space $Pol_1$. Its maps $Pol_1$ to $Pol_3$ and, say, is represented in the monomial bases of $Pol_1$ and $Pol_3$  by the matrix 
$$\begin{pmatrix} \bL_1 & L_{1,2} & L_{1,1} & L_{1,0}\\
                                 0           & \bL_0 & L_{0,1} & L_{0,0}
\end{pmatrix}. 
$$
(Here we used the notation from the proof of Theorem~\ref{th:higHei}.)  Since $r=2$ we need to add to $\dq$ a quadratic  Van Vleck polynomial $V(z)=v_2z^2+v_1z+v_0$  with the undetermined coefficients 
$v_2,v_1,v_0$ which  modifies the above matrix as follows: 
 $$\begin{pmatrix} \bL_1+v_2 & L_{1,2}+v_1& L_{1,1}+v_0 & L_{1,0}\\
                                 0           & \bL_0+v_2 & L_{0,1}+v_1 & L_{0,0}+v_0
\end{pmatrix}. 
$$

The operator $\dq+V(z)$ has a linear Stieltjes polynomial $S(z)=s_1z+s_0$ if and only if  the  vector 
$(s_1,s_0)$ is the left kernel of the latter matrix which leads to the system:
$$\begin{cases} 0=s_1(\bL_{1}+v_2);\\
                              0=s_1(L_{1,2}+v_1)+s_0(\bL_{0}+v_2);\\
                              0=s_1(L_{1,1}+v_0)+s_0(L_{0,1}+v_1);\\
                              0=s_1L_{1,0}+s_0(L_{0,0}+v_0).
\end{cases}$$

Setting $s_1=1$ and $v_2=-\bL_1$ as was explained earlier we get $s_0=\frac{L_{1,2}+v_1}{\bL_1-\bL_0}$ from the $2$-nd equation.  Substituting the obtained variables in the remaining two equations we get the  system of two equations:
$$\begin{cases}
(\bL_1-\bL_0)(v_0+L_{1,1})+(v_{1}+L_{1,1})(v_1+L_{1,2})=0;\\
(v_0+L_{0,0})(v_1+L_{1,2})+(\bL_1-\bL_0)L_{1,0}=0.
\end{cases}
$$
which determines three (not necessarily distinct) pairs $(v_1,v_0)$ which together with $v_2=-\bL_1$ given us three required (not necessarily distinct) quadratic Van Vleck polynomials whose Stieltjes polynomials are of degree exactly $1$.
\end{example}

 Now we finally describe the notion of  natural multiplicity of a given Van Vleck polynomial of an operator $\dq$ used in the introduction.  Let $\widetilde V(z)=\tilde v_rz^r+\tilde v_{r-1}z^{r-1}+...+\tilde v_0$ be some fixed  Van Vleck polynomial of the Heine-Stieltjes problem (\ref{eq:1}), i.e. there exists a (not necessarily unique) polynomial solution $\widetilde S(z)$ of the equation of (\ref{eq:1}) with the chosen $V(z)=\widetilde V(z)$. (Below we use notation from the proof of Proposition~\ref{pr:nonres}.)
 
 \begin{definition}\label{def:multi}
 Given a positive integer $n$ let us define the $n$-th multiplicity $\sharp_n(\widetilde V)$ of $\widetilde V(z)$ as  the usual local algebraic multiplicity of the intersection of the $(r+1)$-dimensional matrix pencil  $\mathcal A_n(v_r,v_{r-1},...,v_0)=A_{\dq,n}+v_rI_0+v_{r-1}I_1+...+v_0I_r$ consisting of 'triangular'  $(n+1)\times (n+r+1)$-matrices  with the set $TM^1\subset TM(n+1,n+r+1)$ of positive corank matrices at the matrix  $A_{\dq,n}+\tilde v_rI_0+\tilde v_{r-1}I_1+...+\tilde v_0I_r.$  Here (as above) $A_{\dq,n}$ denotes the matrix of the action of $\dq$ on the space $Pol_n$ taken  w.r.t monomial basis and  
  $A_{\dq,n}+\tilde v_rI_0+\tilde v_{r-1}I_1+...+\tilde v_0I_r$ is, therefore, the matrix of action of the operator $\dq+\widetilde V(z)$ on $Pol_n$. In case, when $A_{\dq,n}+\tilde v_rI_0+\tilde v_{r-1}I_1+...+\tilde v_0I_r$ does not belong to $TM^1\subset TM(n+1,n+r+1)$, i.e. the operator $\dq+\widetilde V(z)$ does not annihilate any polynomial of degree at most $n$ we set $\sharp_n(\widetilde V)=0.$
 \end{definition} 
 
 \begin{remark} The natural multiplicity of Van Vleck polynomials in Theorems~\ref{th:higHei} and \ref{th:main} while counting those with Stieltjes polynomials of degree at most $n$ is exactly the $n$-th multiplicity from Definition~\ref{def:multi}. 
 
 \end{remark} 
 
 Obviously, for any given Van Vleck polynomial $\widetilde V(z)$ the sequence $\{\sharp_n(\widetilde V)\},\; n=0,1,...$ is a non-decreasing sequence of non-negative integers. Moreover the following stabilization result holds.  
 
 \begin{lemma}\label{lm:multi} For any non-degenerate operator $\dq$ the sequence $\{\sharp_n(\widetilde V)\}$ of multiplicities of any its Van Vleck polynomial $\widetilde V(z)$ stabilizes, i.e there exists $n_{\tilde V}$ such that for all $n> n_{\tilde V}$ one has $\sharp_n(\widetilde V)=\sharp_{n_{\widetilde V}}(\widetilde V).$
  \end{lemma} 
  
 \begin{proof} Indeed, as was mentioned in e.g. the proof of Proposition~\ref{pr:nonres} the leading coefficient $\tilde v_r$ of $\widetilde V(z)$ must necessarily coincide with $-\bL_m$ for  some  non-negative $m$. The sequence $\{|\bL_j|\}$ is strictly increasing starting from some $j_0$, see (\ref{eq:leading}).  Moreover, by Proposition~\ref{pr:nonres} if $\bL_n\neq L_j, j=0,1,...,n-1$ then the total multiplicity of all Van Vleck polynomials whose leading term equals $-\bL_n$ equals $\binom {n+r} {r}$. Therefore, if we take the index value $j_0$ such $|\bL_j|>|\bL_m|$ for all $j\ge j_0$ then the multiplicities 
 $\sharp_j(\widetilde V)$ can not change for $j\ge j_0$ since the total multiplicity increase  is obtained on Van Vleck polynomials with a different leading coefficient when $j$ grows.   
  \end{proof}

\section {Proof of generalized P\'olya's theorems} \label{sec:loc}

 \medskip
   Let us now prove Theorem~\ref{th:higPol} following  
   straightforwardly the recipe of \cite {Po} which in its turn is
   closely related to the proof of the classical Gauss-Lukas theorem.

   \begin{proof}
      Let $(z_{1},\ldots,z_{n})$ denote the set of all roots of a 
        Stieltjes polynomial $S(z)$ of some degree $n$ satisfying the equation
       (\ref{eq:higLame}) with $\al_{i}$'s being complex and $\be_{j}$'s being 
       positive. Then for each $z_{i}$ one has
       $$\frac {
  S^{(k)}(z_{i})}{S^{(k-1)}(z_{i})}+\sum_{j=1}^l\frac{\be_{j}}{z_{i}-\al_{j}}=0.$$
       This equation has the form
\begin{equation}\label{eq:polya}
\sum_{s=1}^p\frac{m_s}{z_{i}-\xi_{s}}+\sum_{j=1}^l\frac{\be_{j}}{z_{i}-\al_{j}}=0,
\end{equation}
       where $(\xi_{1},\ldots,\xi_{p})$ is the set of all roots of
       $S^{(k-1)}(z)$ with $p=n-k+1$ and $(m_1,...,m_p)$ is the set of multiplicities of the roots 
       $(\xi_{1},\ldots,\xi_{p})$.  Notice that by the standard
       Gauss-Lukas theorem all $(\xi_{1},\ldots,\xi_{p})$ lie in the
       convex hull of the set of roots $(z_{1},\ldots,z_{n})$. Assume
       now that the convex hull of $(z_{1},\ldots,z_{n})$ is not
       contained in the convex hull of
       $(\al_{1},\ldots,\al_{l})$. Then there exists some root $z_{i}$
       and an affine line $L\subset \bC$ separating $z_{i}$ from the
       rest of $z_{j}$'s together with all $\al_{j}$'s and $\xi_{m}$'s.
       But then the equation (\ref{eq:polya}) can not hold since all the vectors $z_i-\xi_s$ and $z_i-\al_j$ lie in the same half-plane. 
        \end{proof}

        \begin{rema+} The above argument works for the roots of Van Vleck polynomials $V(z)$ as
        well and  extends to the case $\be_{j}\ge
        0$.
	  \end{rema+}

To settle a much more delicate Theorem~\ref{th:locl} we will prove a number of localization results having an independent interest.

\subsection{Root localization  for Van Vleck and Stieltjes polynomials}Ê

 \begin{definition}\label{def:CT} Given a finite (complex-valued) measure $\mu$ supported on $\bC$ we call by its {\em total mass} the integral $\int_\bC d\mu(\zeta).$ The Cauchy transform $\C_\mu(z)$ of 
 $\mu$  is standardly defined as 
 \begin{equation}\label{eq:CaTr}
 \C_\mu(z)=\int_\bC\frac{d\mu(\zeta)}{z-\zeta}.
 \end{equation}
  \end{definition} 
  
  Obviously, $\C_\mu(z)$ is analytic outside the support of $\mu$ and
has a number of important properties, e.g. that
$\mu=\frac{1}{\pi}\frac{\C_\mu(z)}{\partial \bar z}$ understood in
the distributional sense.  Detailed information about Cauchy transforms
can be found in \cite{Ga}.

\begin{definition}\label{def:CTP} Given a (monic) polynomial $P(z)$ of some degree $m$ we associate with $P(z)$ its {\em root-counting measure} $\mu_P(z)=\frac{1}{m}\sum_j\delta(z-z_j)$ where $\{z_1,...,z_m\}$ stands for the set of all roots of $P(z)$ with repetitions and $\delta(z-z_j)$ is the usual Dirac delta-function supported at $z_j$.   
\end{definition}

Directly from the definition of $\mu_P(z)$  one has that for any given polynomial $P(z)$ of degree $m$ its Cauchy transform is given by $\C_{\mu_P}(z)=\frac{P'(z)}{mP(z)}$.

We start with a rather simple estimate of the absolute value of the Cauchy transform of a probability measure which will help us to prove Proposition~\ref{pr:local}, comp. Lemma~2 in \cite{Ber}. 

\begin{lemma}\label{lm:est}
 Let $\mu$ be a probability measure supported in a disk $D_0$ of radius $R_0$ and centered at $z_0$. Then for any $z$ outside $D_0$ one has the following estimate of the absolute value of its Cauchy transform $\C_\mu(z)$: 
\begin{equation}\label{eq:simpl1}
\frac{1}{|z-z_0-R_0|}\ge |\C_\mu(z)|\ge \frac{1}{2|z-z_0|}.
\end{equation}
\end{lemma}

\begin{proof} The l.h.s. of the above inequality is quite obvious. By (\ref{eq:CaTr}) one has that $|\C_\mu(z)|$ will be maximal if one places the whole unit mass of $\mu$ at the point which has the least distance to $z$ in the admissible support. In our case such a point $p$ is the intersection of  the boundary circle of $D_0$ with the segment $(z,z_0)$. Its distance to $z$ equals $|z-z_0-R_0|$ which gives the required inequality. To settle the r.h.s. let us assume for simplicity that $z_0=0$. Translation invariance of our considerations is obvious. Let us use (\ref{eq:CaTr}) and change the  integration variable as follows: 
 $$\frac{1}{z-\zeta}=\frac{1}{z} \cdot\frac{1}{1-\zeta/z}=\frac{1}{z}\cdot \frac{1}{1-\theta}$$
 where $\theta=\frac{\zeta}{z}$. Since $z$ lies outside $D_0$ and $\zeta$ lies inside $D_0$ one has $|\theta|<1$ which implies  for $w=\frac{1}{1-\theta}$ that one has $\text{Re}(w)\ge \frac 1 2$. Indeed, 
 $$|w-1|=\frac{|\theta|}{|1-\theta|}=|\theta||w|\le |w| \Leftrightarrow |w-1|\le |w| \Leftrightarrow \text{Re}(w)\ge \frac 1 2.$$
Therefore, 
$$|\C_\mu(z)|=\left\vert\int_\bC\frac{d\mu(\zeta)}{z-\zeta}\right\vert=\frac{1}{|z|}\left\vert\int_\bC\frac{d\mu(\zeta)}{1-\theta}\right\vert=\frac{1}{|z|}\left\vert\int_\bC{wd\mu(\zeta)}\right\vert\ge \frac{1}{|z|}\left\vert\int_\bC{\text{Re}(w)d\mu(\zeta)}\right\vert\ge\frac{1}{2|z|}.$$ 
\end{proof} 

Using Lemma~\ref{lm:est} we now settle Proposition~\ref{pr:local} (see Introduction). 

\begin{proof} Take a pair $(V(z), S(z))$ where $V(z)$ is some Van Vleck polynomial and $S(z)$ is its  corresponding Stieltjes polynomial of degree $n$.  Let $\xi$ be the  root of either $V(z)$ or  $S(z)$  which has the maximal modulus among all roots of the chosen $V(z)$ and $S(z)$. We want to show that there exists a radius $R>0$ such that 
$|\xi|\le R$ for any $\xi$ as above and as soon as $n$ is large enough. Substituting $V(z), S(z), \xi$ in (\ref{eq:1}) 
and using (\ref{eq:BasicOp}) we get the relation:  
$$
Q_k(\xi)S^{(k)}(\xi)+Q_{k-1}(\xi)S^{(k-1)}(\xi)+...+Q_1(\xi)S'(\xi)=0,
$$
dividing which by its first term we obtain:
\begin{equation}\label{eq:inter}
1+\sum_{j=1}^{k-1}\frac{Q_j(\xi)S^{(i)}(\xi)}{Q_k(\xi)S^{(k)}(\xi)}=0.
\end{equation}
Notice that the rational function $b_i(z):=\frac{S^{(i+1)}(z)}{(n-i)S^{(i)}(z)}$ is the Cauchy transform of the polynomial $S^{(i)}(z)$.  Easy arithmetic shows that 
$$S^{(i)}(z)=\frac{S^{(k)}(z)}{(n-k+1)...(n-i)\prod_{j=i}^{k-1}b_j(z)}\Leftrightarrow \frac{S^{(i)}(z)}{S^{(k)}(z)}=\frac{(n-k)!}{(n-i)!\prod_{j=i}^{k-1}b_j(z)}.$$ 
Notice additionally, that by the usual Gauss-Lucas theorem all roots of any $S^{(i)}(z)$ lie within the convex hull of the set of roots of $S(z)$. In particular, all these roots lie within the disk of radius $|\xi|.$ 
Therefore, using Lemma~\ref{lm:est}  we get 
\begin{equation}\label{eq:imp}
\left |\frac{Q_i(\xi)S^{(i)}(\xi)}{Q_k(\xi)S^{(k)}(\xi)}\right | \le \frac{|Q_i(\xi)|}{|Q_k(\xi)|}\frac{(n-k)!}{(n-i)!}2^{k-i}|\xi|^{k-i}.
\end{equation}
Notice that since $Q_k(z)$ is a monic polynomial of degree $k+r$ (recall that $r$ is the Fuchs index of the operator $\dq$) then one can choose a radius $R$ such that for any $z$ with $|z|>R$ one has 
$|Q_k(z)|\ge \frac{|z|^{k+r}}{2}$. Now since for any $i=1,...,k-1$ one has $\deg Q_i(z)\le i+r$ we can choose  a positive constant $K$ such that $|Q_i(z)|\le K |z|^{i+r}$ for all $i=1,...,k-1$ and $|z|>R.$ We want to show that $\xi$ can not be too large for a sufficiently large $n$. Using our previous  assumptions and assuming additionally that $|\xi|>R$ we get  
$$\left |\frac{Q_i(\xi)S^{(i)}(\xi)}{Q_k(\xi)S^{(k)}(\xi)}\right | \le \frac{|Q_i(\xi)|}{|Q_k(\xi)|}\frac{(n-k)!}{(n-i)!}2^{k-i}|\xi|^{k-i}\le \frac{K\cdot 2^{k-i+1}}{(n-i)...(n-k+1)}.$$
Now we can finally choose $N_0$ large enough such that for all $n\ge N_0$, all $i=1,...,k-1$ and any $|\xi|>R$ one has that 
$$\left |\frac{Q_i(\xi)S^{(i)}(\xi)}{Q_k(\xi)S^{(k)}(\xi)}\right | \le \frac{K\cdot 2^{k-i+1}}{(n-i)...(n-k+1)}<\frac{1}{k-1}.$$
But then obviously the relation (\ref {eq:inter}) can not hold for all $n\ge N_0$ and any $|\xi|>R$ since 
$$\left|\sum_{j=1}^{k-1}\frac{Q_j(\xi)S^{(i)}(\xi)}{Q_k(\xi)S^{(k)}(\xi)}\right|\le \sum_{j=1}^{k-1}\left|\frac{Q_j(\xi)S^{(i)}(\xi)}{Q_k(\xi)S^{(k)}(\xi)}\right|<\sum_{i=1}^{k-1}\frac{1}{k-1}<1.$$ 
\end{proof}Ê

Now we will strengthen the arguments in the proof of   Proposition~\ref{pr:local} in order to settle Theorem~\ref{th:locl}.  Denote by $\mathcal  R_{Q_k}$ the maximal distance between the origin and  $Conv_{Q_k}$. The following statement holds. 

\begin{lemma}\label{lm:newest} For any non-degenerate higher Lam\'e operator $\dq$ and a given number $\delta>0$ there exists a positive integer $N_\delta$ such that the roots of all Van Vleck polynomials $V(z)$ possessing a  Stieltjes polynomial $S(z)$ of degree $\ge N_\delta$ as well as the roots of these Stieltjes polynomials lie in the disk $|z|\le \mathcal  R_{Q_k}+\delta$.
\end{lemma}

\begin{proof}ÊNotice that once $\delta$ is fixed  the quotient $\frac{|Q_i(z)|}{|Q_k(z)|}$ is bounded from above for each $i=1,...,k-1$ if we assume that $|z|\ge  \mathcal R_{Q_k}+\delta$. Indeed, all the roots of $Q_k(z)$ lie within  the disk of radius $\mathcal R_{Q_k}$ centered at the origin and each $Q_i(z)$ has a smaller degree than $Q_k(z)$. Consider now again the estimate \eqref{eq:imp}. Since we now know that 
$\xi$ lies in some bounded domain for all possible polynomials $V(z)$ and $S(z)$ of sufficiently high degree and that the quotient $\frac{|Q_i(z)|}{|Q_k(z)|}$ is bounded from above outside the disk of radius 
$\mathcal  R_{Q_k}+\delta$ we get that the right-hand side of \eqref{eq:imp} goes to $0$ when $n\to\infty$ under the assumption that $\xi$ stays outside the latter disk. Looking again at \eqref{eq:inter} we see that by the latter argument it can not hold for  $|\xi|\ge \mathcal  R_{Q_k}+\delta$ when $n\to \infty$. This contradiction proves the lemma.
\end{proof}Ê

To finish the proof of Theorem~\ref{th:locl} notice that the choice of the origin is in our hands, i.e we can make an arbitrary affine shift of the independent variable $z$ and use the same arguments. Since the convex hull $Conv_{Q_k}$ is the intersection of all disks centered at different points and containing $Conv_{Q_k}$ we can for any chosen $\eps>0$ find the intersection $\mathcal K$ of finitely many disks in $\bC$ such that $\mathcal K$ contains 
$Conv_{Q_k}$ but is contained in $Conv_{Q_k}^\eps$.  (One can choose one such disk for each edge  of the boundary of $Conv_{Q_k}$ putting its center sufficiently far away on the line perpendicular to the edge and passing   through its middle point.) Then  
since $\mathcal K$ is the intersection of finitely many disks we can  applying Lemma~\ref{lm:newest}  find such $N_\eps$ that  all roots of all $V(z)$ and $S(z)$ for all $n\ge N_{\eps}$ lie in $\mathcal K$. \qed

  \section {'On the existence and number of Lam\'e functions of
higher degree', by E.~Heine }\label{sec:heine} 

\subsection{ Comments on Heine's result and  history around it.}  
        
        As   Heine himself mentions in  \cite{He}  the requirement of algebraic independence of the coefficients of $Q_2(z)$ and $Q_1(z)$ being sufficient  is too strong and restrictive for his purposes but he fails to give any other explicit condition guaranteeing the same result, see Theorem~\ref{th:Heine}.  Heine's original motivation for the consideration (\ref{eq:comLame}) comes from the classical Lam\'e equation (\ref{eq:standLame}) in which case $Q_1(z)$ and $Q_2(z)$ are very much algebraically dependent, namely, $Q_1(z)=Q'_2(z)/2$. Also in order to prove that the upper bound ${n+l-2\choose n}$ is actually achieved for algebraically independent $Q_1(z)$ and $Q_2(z)$ Heine uses an inductive argument where he forces $Q_2(z)$ and $Q_1(z)$ to become algebraically dependent in a special way.  Theorem~\ref{th:HSV}  is another clear indication that the algebraic independence is apparently an inappropriate condition for the goal. 
       
          Interpretation of Heine's text written in a rather cumbersome
       19-th century German and the exact statements it contains seems to
       create  difficulties for mathematicians starting from 1870's and  up to now, see e.g.
       \cite {Mar}.  The main  classical sources, namely, \cite{St} from 1885, \cite {Po}  from 1912 and \cite {Sz}  from 1939 are  not too clear about what it is that Heine actually  proved and under what assumptions  on $Q_2(z)$ and $Q_1(z)$ one can guarantee that for a given  positive $n$ the number of possible polynomial pairs $(V,S)$ such that the corresponding $S$ has degree exactly $n$ is finite and bounded by  ${n+l-2\choose n}$. Being aware  of the existence of a  gap in his proof Heine seems to  covers  by a reference to   a letter of his friend Leopold Kronecker who has  (under unknown conditions)  shown that for a given degree $n$ the eliminant of the system of algebraic equations  defining the coefficients of  the polynomial $V(z)$ does not vanish identically.  This statement is equivalent to the finiteness of the number of these polynomials.  Heine mentions also a 
   short note of  Kronecker's on this topic presented  in the January issue of Monatsbericht der Berliner Akademie from 1864.  Unfortunately the track ends here. All one can find in this issue is the phrase that at  the meeting on the section of physics and mathematics of the Prussian Academy of Science held on November 14, 1864 "Herr Kronecker las \"uber die verschieden Faktoren des Discriminante von Eliminantions-Gleichungen ",  i.e. "Herr Kronecker gave a lecture on different factors of the discriminant of the elimination equation."     In attempt  to overbridge this gap we undertook the  task
       of translating and (even more so) decoding Heine's arguments. 
  In what follows we give a frase-by-frase translation of portions of \S 135 and \S
136 from Heine's book \cite{He} which are relevant for our consideration. Some of the phrases were
difficult to understand literally and we provided our interpretation and comments 
of the  content placed within  the slash signs. We allowed ourselves to correct several obvious misprints without a special mention,   tried  to keep (to certain extent) the
language flavor and preserved the enumeration of the formulas of the original text. 

\subsection {Translation}

\S 135 ...

Next we ask which conditions the polynomials $\chi(x)$ and
$\theta(x)$ must satisfy in order for the differential equation
$$\psi(x)\frac {d^2W}{du^2} +\chi(x)\frac
{dW}{du}+\theta(x)W=0\phantom{XXXXXX}(88)$$
    to have a solution which is a polynomial of degree $n$ in
    $x$ assuming that $\psi$ is of degree $p+1$ and $\chi$ and $\theta$
    are of degrees at most $p$ resp. $p-1$. 
    \noindent 
    /In fact, $\chi(x)$ is fixed and $\theta(x)$ is a variable polynomial. It is 
not clear why Heine talks
    about both $\chi(x)$ and $\theta(x)$ here./

   This is always the case when
    it is the question about, as in the case of  Lam\'e functions, a 
    differential equation whose general solution does not contain any
    higher transcendentals than a rational function of integrals of
    algebraic functions and which has a definite order at $x=\infty$. 
    /This and the next two phrases explain why one should assume that  $\deg(\chi)<\deg(\psi)$./

  We say about a function $W$ that it is
    of the order $\alpha$ for the finite value $x=a$, if
$(x-a)^{\alpha-\epsilon}W$ and
    $(x-a)^{\alpha+\epsilon}W$ will be $0$ respectively $\infty$ for $x=a$,
    however small $\epsilon$ is taken; we assign to it  the
    order $\alpha$  at infinity, if $x^{-\alpha-\epsilon}W$ and
$x^{-\alpha+\epsilon}W$
    for $x=\infty$ becomes $0$ resp. $\infty$.  Thus, for example, $\log
    x$ has a definite order, namely $0$.

    If $y$ and $z$ are two particular solutions of (88) then one has
    $$\log  (yz'-zy')=\int\frac {\chi}{\psi}dx.$$

  If $\chi$ were not of smaller degree than $\psi$ then $yz'-zy'$ would
    at $x=\infty$ go to $0$ or $\infty$ as an exponential function  and
    would therefore have no order. 

   For the solution to have an order for every $x$ where $\psi(x)$
    vanishes, $\frac {\chi(x)}{\psi(x)}$, after possible cancellations 
     can in the denominator only have different factors.
    This follows from the same equality between two particular solutions
    which we used above. 

  The following theorem answers the question posed at the
    begining:

\medskip
    {\em If the two polynomials $\psi(x)$
    and $\chi(x)$ are given, the first of degree $p+1$, the second of
    degree $p $, then for exactly $\frac
    {(n+1)(n+2)\ldots(n+p-1)}{1.2\ldots(p-1)}$ different functions $\theta(x)$,
    there exists a particular solution of (88) which is a polynomial of
    degree $n$ in $x$. } 

  \medskip
  For $p=1$ we understand the number given above which in general
      might be denoted by $(n,p)$ as $1$. It is assumed above that the 
coefficients of
$\psi$ and $\chi$ are mutually
      independent. I call the numbers $a,b,$ etc. mutually independent
      when there is no algebraic equation with integer coefficients
      which they satisfy. 
         /As stated above Heine actually proves his
      theorem under the assumption of the algebraic independence of
      the coefficients./

     It may be  immediately  added here that we will say about the
      numbers $a,b,\ldots$ which  already satisfy one or more algebraic
      equations  that "they are further specialized" when
      in addition to these equations they satisfy one or more additional
      equations - which of course are not allowed to contradict the
      earlier ones.

          The just mentioned assumption for the validity of the theorem
      demands more than is necessary. One could say with the same right
      that a polynomial of degree $n$ in $x$ with mutually
      independent coefficients vanishes for $n$ different values of $x$
      while for this it would suffice that the coefficients did not
      satisfy a particular equation  with integer coefficients,
      namely, the well-known one which describes the coincidence of
      roots.  Also for the validity of our theorem it suffices that the
      coefficients do not satisfy  certain finite number of algebraic
      equations, which equations in every case can be found but not in
      a comprehensible way.  
    /As  we mentioned in the introduction Heine realizes  that for a given 
fixed $n$ the set of all pairs of polynomials $(\psi(x),\chi(x))$ of degrees at most $p+1$ and $p$ respectively   for which there are less   than $(n,p)$ functions $\theta(x)$ solving 
      the problem under consideration  is an algebraic hypersurface with integer coefficients. It is by no means clear to us  how Heine could possibly conclude that all the coefficients of the discriminantal equation are integers in the basis of the coefficients of $\psi(x)$ and $\chi(x)$. The development of the corresponding theory can be traced back to Cayley, see appendix in \cite {GKZ}, but no general results  
      were obtained until much later./

       \S 136. To obtain the proof of the theorem one substitutes in (88)
      polynomials of degree $n$ and $p-1$ for $W$ and $\theta$, namely
      $$W=x^n+g_{1}x^{n-1}+\ldots+,$$
      $$\theta=k_{0}x^{p-1}+k_{1}x^{p-2}+k_{2}x^{p-3}+\ldots$$

     It is clear that the necessary and sufficient condition that $W$
      satisfies the equation (88) is that certain $n+p$ equations are
      satisfies which are linear in both the $g_{i}$'s and $k_{j}$'s and in
      the coefficients of $\psi$ and $\chi$.
       To show the structure of these,
      without having to work with too clumsy formulas
      I present them for the case $p=3$.

  Let the given functions be
      $$\psi(x)=x(c_{0}x^3+c_{1}x^2+c_{2}x+c_{3}),$$
      $$\chi(x)=b_{0}x^3+b_{1}x^2+b_{2}x+b_{3}$$
      and the sought functions
      $$W(x)=g_{0}x^n+g_{1}x^{n-1}+g_{2}x^{n-2}+\ldots +g_{n},$$
      $$\theta(x)=k_{0}x^2+k_{1}x+k_{2}.$$
      /Notice that Heine apparently realizes that, in general, it might be impossible to find $W(x)$ as a polynomial of degree exactly $n$ and introduces even the leading coefficient as a new variable without special explanations./ 

  For $W$ to satisfy the differential equation (88) the
      coefficients $g_{i}$'s and $k_{j}$'s must satisfy the system of equations:

     \begin{align*}
     0 = & g_{0}[k_{0}+nb_{0}+n(n-1)c_{0}]; \\
     0 = &
g_{1}[k_{0}+(n-1)b_{0}+(n-1)(n-2)c_{0}]+g_{0}[k_{1}+nb_{1}+n(n-1)c_{1}];\\
     0 = &
 g_{2}[k_{0}+(n-2)b_{0}+(n-2)(n-3)c_{0}]+g_{1}[k_{1}+(n-1)b_{1}+(n-1)(n-2)c_{1}]+\\
     &+g_{0}[k_{2}+nb_{2}+n(n-1)c_{2}];\\
     0 = &
 g_{3}[k_{0}+(n-3)b_{0}+(n-3)(n-4)c_{0}]+g_{2}[k_{1}+(n-2)b_{1}+(n-2)(n-3)c_{1}]+\\
     &+g_{1}[k_{2}+(n-1)b_{2}+
     (n-1)(n-2)c_{2}]+g_{0}[nb_{3}+n(n-1)c_{3}];\\
      0 = &
 g_{4}[k_{0}+(n-4)b_{0}+(n-4)(n-5)c_{0}]+g_{3}[k_{1}+(n-3)b_{1}+(n-3)(n-4)c_{1}]+\\
     &+g_{2}[k_{2}+(n-2)b_{2}+
     (n-2)(n-3)c_{2}]+g_{1}[(n-1)b_{3}+(n-1)(n-2)c_{3}];\\
     \ldots
     \end{align*}

  In this way the equations will continue to be formed so that the
     next one will give the relation between the four $g_{i}$'s with the
     indices $5,4,3,2$. The final equations will be

       \begin{align*}
     0 =&  g_{n-1}[k_{0}+b_{0}]& + g_{n-2}[k_{1}+2b_{1}+2\cdot 1 c_{1}]&
     + g_{n-3}[k_{2}+3b_{1}+3\cdot 2 c_{2}]& + g_{n-4}[4b_{3}+4\cdot 3
c_{3}]&; \\
     0 =&          g_{n}[k_{0}]& + g_{n-1}[k_{1}+1\cdot b_{1}]&
     + g_{n-2}[k_{2}+2b_{2}+2\cdot 1 c_{2}]& + g_{n-3} [3b_{3}+3\cdot 2
c_{3}]&;\\
     0 =& & g_{n}[k_{0}]& + g_{n-1}[k_{1}+1\cdot b_{1}]& +
g_{n-2}[2b_{3}+2\cdot 1 c_{3}]&;\\
     0 =& & &   g_{n}[k_{0}]& + g_{n-1}[b_{3}].&
     \end{align*}

   From the first equation the coefficient $k_{0}$ is completely determined in
    terms of the given
     coefficients $b_{0}$ and $c_{0}$ of $\psi$ and $\chi$; the
     next $n$
     equations give all $g_{i}$'s expressed in terms of the same known
coefficients
     $b_{i}$'s and $c_{j}$'s and the $(p-1)$ (in our example $2$) unknowns
     $k_{1},k_{2},\ldots$.
    The values of $g_{i}$'s that are obtained from the
     second to the $(n+1)$st equations when substituted in the last
     $(p-1)$ equations, will then give $(p-1)$ equations of higher
     degrees
     between the unknowns $k_{1},k_{2},\ldots,k_{p-1}$ and the known
     coefficients of $\psi$ and $\chi$, which only appear rationally
     in these equations. 
        /Heine apparently means that $g_{i}$'s will be given by rational
     functions of $c_{l}$'s and $b_{m}$'s and substituting these one
     gets a system of rational equations defining $k_{j}$'s./

Once the $k_{j}$'s have been
     determined from these $p-1$ equations, the substitution of the found
     values in the 2nd to the $(n+1)$st equations will give the $g_{i}$'s.
Two systems of related $k_{j}$'s are called {\em different}, i.e. 
     the two systems $k_{1},\ldots,k_{p-1}$ and $k'_{1},\ldots,k'_{p-1}$
     are called {\em different} when they are not equal.
     One realizes with the
     full confidence from the form of the $2$nd to the $(n+1)$-st equation
     that every system $k_{j}$'s corresponds to a system of $g_{i}$'s and
     different systems of $k_{j}$'s correspond to different system of
     $g_{i}$'s. 
         /The latter statement of Heine is false as is. It requires that the diagonal entries in the uppertriangular system are non-vanishing (see our nonresonance condition in the introduction). But it is certainly true if the coefficients of $\psi(x)$
    and $\chi(x)$ are algebraically independent./

     One obtains thus that

\noindent
    {\em  There exist as many different equations (88) and
     therefore as many different polynomials $W$ of degree $n$ as there
     are different systems of $k_{j}$'s.}

     Next it is realized that

\noindent 
     {\em The degree of the elimination equation is at
     most $(n,p)$, that is there can only be at most $(n,p)$ different
     systems of $k_{j}$'s.}

    If one throws a glance at the $(n+p)$ equations,
     which, with the exception of the first one for $k_{0}$,  one can
     find above for the special case that $p=3$, one will perhaps not
     realize the truth of this statement immediately, and instead believe
     that the degree of the elimination equation is larger. /One is supposed to disregard the $1$st equation in the system above and study the remaining $(n+p)$ equations./ But if one
     instead of $k_{2},k_{3},\ldots,$ substitutes
     $x_{2}^2,x_{3}^3,\ldots$ where the lower numbers are indices and
     the upper numbers are exponents, and for symmetry we use
     $x_{i}\leftrightarrow k_{i}$, one realizes immediately that
     $g_{1},g_{2},\ldots,g_{n}$ are polynomials in the $x_{i}$'s of degree
     $1,2,\ldots,n$ resp. so that after the substitution the $(p-1)$ last
     equations will have degrees $n+1,n+2,\ldots,n+p+1$ in $x_{i}$'s. That means
     that the degree of the elimination equation will at most grow to
     $(n+1)(n+2)\ldots(n+p+1)$.
  If one  takes into account that every value of
     $k_{1},k_{2},\ldots$ corresponds to one value of $x_{1}$, two values
     of  $x_{2}$, three values of $x_{3}$ then the above assertion is
     proved. 
         /This is an excellent passage! What Heine does is called in the
     modern language of algebraic geometry the weighted Bezout theorem, see e.g. \cite {Do}. The author  was unable to find a reliable  proof of this result in the literature prior to 1970./

\noindent 
     {\em Under the assumptions that the elimination equation is not
    identically zero it will indeed have the above mentioned degree and
    give $(n,p)$ different systems of $k$'s.}
    
    /Crucial claim but not completely proved below./ 

     There exist as I will show below indeed $(n,p)$ distinct systems of
    the coefficients if the coefficients of $\psi$ and $\chi$ are
    specialized in a certain way. 
    /Heine will show by induction that for a special choice of
    $\psi$ and $\chi$ one can obtain exactly $(n,p)$ distinct solutions. But instead 
    of algebraically independent coefficients of $\psi$ and $\chi$ he needs to make them dependent to get an example of $(n,p)$ distinct  solutions. This is correct as soon as one knows that even for the specialized situation the total number of solutions is finite. This finiteness 
    probably follows from his specific choice of specialization, see below./

That this elimination resultant does not vanish identically
    follows from the next observation which I take from a letter of my
    friend Kronecker. 
       /One needs to check also that the system of equations has (under
   very unclear non-degeneracy assumptions on  the coefficients
    of $\psi$ and $\chi$) only isolated solutions. This is equivalent to the non-vanishing of the eliminant./

   If in the mentioned final equation which will determine the
    functions $\theta(x)$ and $W(x)$ all coefficients vanish then
    according to the general principle of elimination will at least one
    of the roots of $W(x)=0$ be unrestricted. 
       /This is an interesting although a rather obvious  observation. Notice that the elimination theory hardly at all existed in 1870's./ 

  If one assigns to this root all values for which $\psi(x)$
    vanishes then one gets through this (procedure) certain restrictions
    on the function $\chi(x)$ but the latter function does not satisfy them even after the mentioned specialization. 

   Herr Kronecker added in the mentioned message that these
    restrictions are actually satisfied and one of the roots of $W(x)=0$ remains
    undetermined if both $\psi(x)$ and $\chi(x)$ have the properties
    that for the known function $\theta(x)$ both solutions  of (88) 
are polynomials in $x$.
\footnote{
    In Monatsbericht der Berliner Akademie from January 1864 added
    (ackomplished) Herr Kronecker my message with the following

    Introducing the roots of $W(x)=0$ as variables in the equations
    $$\psi(x_{k})W''(x_{k})+\chi(x_{k})W'(x_{k})=0,\; k=1,2,\ldots,n$$
    which define them and substituting the coefficients of $W'$ and $W''$
    through the symmetric functions of $x_{1},\ldots,x_{n}$ one sees
    directly that one of the unknowns $x$ remains arbitrary if the
    elimination equation vanishes.
Through a simple transformation of this system of equations one can
    determine the degree of the final equation and at the same time
    prove that certain coefficients are different from $0$ as long as
    there are no special conditions on the functions $\psi$ and
    $\chi$. 
       /A rathe unclear  proof of the nonvanishing of the eliminant./}


    Concerning the number of systems with specialized $\psi$ and
    $\chi$ I set such relations between the coefficients which give
    that $\psi$ has a factor $(x-a)$ twice and $\chi$ has it once. 
    /Then the algebraic independence is lost here since $\psi$ has a double root and therefore lies on a   discriminantal  surface./ 
  Then all $W$ satisfying (88) have the form
    $$U(n); (x-a)U(n-1);\ldots; (x-a)^nU(0),$$
    where $U$ are as in \S 123 polynomials coprime with $(x-a)$ are
    their degrees are given in parenthesis to the left of the letter
    $U$.
  Under the substitution of this expression in (88) one gets 
    for every $U$ an equation like (88) in which instead of $\psi$ and
    $\chi$ appear polynomials with unrestricted coefficients which are
    not of degree $p+1$ and $p$ but instead of $p$ and $p-1$. /Apparently the logics 
    here is as follows. We can use our result to prove that the number $(n,p)$ of simple solutions is obtained if we can show that for the above specialization the total number of solutions 
    is finite./ 

     If one assumes that the general statement which we are proving is
    settled if $\psi$ is the product of $p$ linear factors (and for the
    product of 2 linear factors this is easy to show) then one gets the
    situation when $\psi$ consists of $p+1$ factors of which two are
    coinciding, alltogether
    $$(n,p-1)+(n-1,p-1)+(n-2,p-1)+\ldots+ (0,p-1)$$
    which after summation gives $(n,p)$ different $W$, i.e. $(n,p)$
     different $\theta$ just as many as different systems of
     $k$'s.  /This accomplished  the induction step. What one misses is mentioning  that the total number of solutions for Heine's specialization is finite.  But this follows from his representation of all  solutions as 
     $U(n); (x-a)U(n-1);\ldots; (x-a)^nU(0),$. In each of these cases we already  know that the number of solutions is finite and all of them are simple. To be completely rigorous one should  use a double induction on $p$ and $n$ as we did in \S~\ref{sec:proofs}. Additional simplification of the order $2$ case compared to the general order $k$ case considered above comes from the fact that during the above  specialization one can assume that the polynomials $\widetilde \psi$ and $\widetilde \chi$ such that 
     $\psi=(x-a)\widetilde \psi$ and $\chi=(x-a)\widetilde \chi$ have algebraically independent coefficients which does not work for higher order case./

\medskip 
\section {Final Remarks}
\label{sec:final}
Let us formulate a number of relevant questions and conjectures. 
\medskip

\begin{problem} 
Is it possible to describe when a linear ordinary differential equation with polynomial coefficients admits    at least 2 polynomial solutions? 
\end{problem}Ê

The prototype result  of Varchenko-Scherbak gives a satisfactory answer for equations of the second order. The answer to the latter question allows to detect the appearance of multi-dimensional families of Stieltjes polynomials. 

\begin{problem} 
Under the nonresonance assumption (\ref{eq:nonres}) is it possible to obtain explicitly  the discriminantal surface which shows when a Van Vleck polynomial attains a non-trivial multiplicity. 
\end{problem}Ê

This question addresses the problem of explicit determination of the discriminantal surface mentioned in Heine's proof. Some discussion of this problem can be found in \cite{SS}. 

\begin{problem} Explain how  the number of Van Vlecks polynomials having Stieltjes polynomials for a certain given degree $n$ can drop below $\binom {n+r}{r}$?  
\end{problem}

The next question addresses the issue of location of the roots of Van Vleck and Stieltjes polynomials. 

\begin{problem} Under what assumptions on $\dq$ the roots of any its Van Vleck and Stieltjes polynomials  lie in the convex hull of its leading coefficient $Q_k(z)$? 

The basic examples are provided by Stieltjes's and Polya's theorems.   
\end{problem}Ê

Finally, 

\begin{problem} Is it possible to extend the results of this paper to the case of degenerate higher Lam\'e operators?   
 
 T.~Bergkvist  \cite {Ber}Ê has obtained a number of interesting results and conjectures in the case of degenerate exactly solvable operators.  Motivated by her resultsÊ we formulate the following conjecture. 
    
    \begin{conjecture}   For any degenerate Lam\'e operator and any positive integer $N_0$ the  union of all the roots to polynomials $V$ and $S$ taken over $\deg S\ge N_0$ is always unbounded. Therefore, this property is a key distinction between non-degenerate and degenerate Lam\'e operators.   
   \end{conjecture}

 \end{problem}

\end{document}